\documentclass[aps,prd,nofootinbib,showpacs,floatfix,twocolumn]{revtex4}  
\usepackage{bm}
\usepackage{latexsym}
\usepackage{dcolumn}
\usepackage{amsfonts,amssymb,amsmath}
\usepackage{graphicx,epsfig}
\usepackage{psfrag}
\usepackage{subfigure}
\usepackage{feynmf}
\usepackage{rotating}
\usepackage{hyperref}
\usepackage{enumitem}
\usepackage{color}
\usepackage{makecell}
\usepackage[normalem]{ulem}
\usepackage[export]{adjustbox}
\hypersetup{
    bookmarks=true,         % show bookmarks bar?
    unicode=false,          % non-Latin characters in Acrobat’s bookmarks
    pdftoolbar=true,        % show Acrobat’s toolbar?
    pdfmenubar=true,        % show Acrobat’s menu?
    pdffitwindow=false,     % window fit to page when opened
    pdfstartview={FitH},    % fits the width of the page to the window
    pdftitle={My title},    % title
    pdfauthor={Author},     % author
    pdfsubject={Subject},   % subject of the document
    pdfcreator={Creator},   % creator of the document
    pdfproducer={Producer}, % producer of the document
    pdfkeywords={keyword1} {key2} {key3}, % list of keywords
    pdfnewwindow=true,      % links in new window
    colorlinks=true,       % false: boxed links; true: colored links
    linkcolor=blue,          % color of internal links
    citecolor=cyan,        % color of links to bibliography
    filecolor=magenta,      % color of file links
    urlcolor=blue,           % color of external links
    linktocpage=true
}

\def\beq{\begin{equation}}
\def\eeq{\end{equation}}
\def\br{\begin{eqnarray}}
\def\er{\end{eqnarray}}
\def\benu{\begin{enumerate}}
\def\efnu{\end{enumerate}}

\def\l{\left}
\def\r{\right}

\usepackage{soul}

\begin{document}

\title{Dark Twilight Joined with the Light of Dawn to Unveil the Reionization History}
\author{Daniela Paoletti} \email{daniela.paoletti@inaf.it}
\affiliation{INAF OAS Bologna, Osservatorio di Astrofisica e Scienza dello Spazio di Bologna, via Gobetti 101, I-40129 Bologna, Italy  \\INFN, Sezione di Bologna, via Irnerio 46, 40126 Bologna, Italy}
\author{Dhiraj Kumar Hazra} \email{dhiraj@imsc.res.in}
\affiliation{The  Institute  of  Mathematical  Sciences,  HBNI,  CIT  Campus, Chennai  600113,  India\\ INAF OAS Bologna, Osservatorio di Astrofisica e Scienza dello Spazio di Bologna, via Gobetti 101, I-40129 Bologna, Italy}
\author{Fabio Finelli} \email{fabio.finelli@inaf.it}
\affiliation{INAF OAS Bologna, Osservatorio di Astrofisica e Scienza dello Spazio di Bologna, via Gobetti 101, I-40129 Bologna, Italy\\INFN, Sezione di Bologna, via Irnerio 46, 40126 Bologna, Italy}
\author{George F. Smoot}\email{gfsmoot@lbl.gov}
\affiliation{IAS TT \& WF Chao Foundation Professor, IAS, Hong Kong University of Science and Technology, Clear Water Bay, Kowloon, 999077 Hong Kong, China. \\
Paris Centre for Cosmological Physics, Universit\'{e} de Paris, emertius, CNRS,  Astroparticule et Cosmologie, F-75013 Paris, France A, 10 rue Alice Domon et Leonie Duquet,75205 Paris CEDEX 13, France. \\
Donostia International Physics Center (DIPC), 20018 Donostia, The Basque Country, Spain. \\
Energetic Cosmos Laboratory, Nazarbayev University, Astana, Kazakhstan}
\date{\today}

\begin{abstract}
Improved measurement of the Cosmic Microwave Background polarization from Planck      
allows a detailed study of reionization beyond the average optical depth.
The lower value of the optical depth disfavours an early onset and an early completion of reionization in
favour of a redsfhit range where different astrophysical probes provide sensible information on the sources of 
reionization and the status of the intergalactic medium.                                  
In this work we extend our previous study in which we constrained reionization by 
combining three different probes - CMB, UV luminosity
density and neutral hydrogen fraction data~\cite{PRL} - in both treatment and data: we first allow variation in the UV source term varying the product of the efficiency of conversion of UV luminosity into ionizing  photons and the escape fraction
together with the reionization and cosmological parameters, and then we investigate the impact of a less
conservative cut for the UV luminosity function.
We find that the estimate for the efficiency is consistent within 95\% C.L. with the fixed value 
we considered in our previous results and is mostly constrained by QHII data.
We find that allowing the efficiency to vary does not affect significantly our results for the average optical depth  for monotonic reionization histories, recovering $\tau=0.0519_{-0.0008}^{+0.0010}$ at 68 \%CL, consistent with our previous studies.
Using a less conservative cut for the UV luminosity function, we find 
$\tau=0.0541_{-0.0016}^{+0.0013}$ at 68 \% CL, due to the faint end of the 
luminosity function in the data we use, that also prefers a larger contribution from higher redshifts. 
\end{abstract}

\maketitle

\section{Introduction}
The epoch of reionization in a sense represents the dawn of the present Universe that closed the dark ages descended after recombination. Some half billion years after recombination the first sources of light, and ionizing radiation, lit up reionizing the Universe. During this epoch the Universe underwent one of its most important phase transitions and the study of its dynamic and the quest for its sources are some of the most active fields in cosmology especially in perspective of future experiments.
From an observational point of view, reionization presents multiple aspects. Probes of the Inter Galactic Medium (IGM) status are represented by bright sources at high redshifts as Quasars, and in particular the observations of the Lyman series, and Gamma Ray Burst that provide a snapshot of the IGM around the source \citep{Fan2006:annurev,Fan2006,Mortlock2011,Bolton2011,McGreer:2014qwa, Totani:2005ng,McQuinn:2007gm, Schroeder2013,Greig:2016vpu,Davies:2018pdw,McQuinn:2007dy,Ouchi2010,Ono2012,Caruana2013,Schenker2014,Tilvi2014,Pentericci2014,Sobacchi2015,Mason:2017eqr,Mason:2019ixe}. The Cosmic Microwave Background (CMB) represents instead a probe which is sensitive to the integrated history of reionization and is the only probe capable of reaching very high redshifts to test the onset of reionization. Finally, the improvements of observational facilities has allowed tracing
the source of the ionizing radiation, in the form of the UV luminosity density coming from star-forming galaxies \cite{Ishigaki2018}. 

Beyond its astrophysical importance per se, reionization plays a crucial role in cosmology. The integrated optical depth is one of the six parameters of the standard cosmological model and after seventeen years from its first determination in 2003~\cite{Spergel:2003cb} is still the parameter with the largest uncertainty in the standard cosmological model and the only one over the percent level accuracy of the Planck 2018 release~\cite{Planck2018:param}. 
This uncertainty is mainly related to the difficulty of the measurement with the CMB. 
CMB is mainly sensitive to the optical depth in the E-mode polarization on large angular scales. 
In particular the reionization bump for $\ell<20$ is sensitive to the value of the optical depth and the duration of reionization, 
but it lies in a region of the multipole space where cosmic variance, systematics and foreground contamination strongly affect the measurements. 
Planck 2018 data are the state of art of the E-mode polarization  at large angular scales and their improvement has  halved the uncertainty on the optical depth 
and for the first time reconciles the determination from the CMB with the ones coming from astrophysical sources~\citep{Planck2018:param,Akrami:2020bpw,Pagano:2019tci,Delouis:2019bub}.
Using this combination, an early onset of reionization  is now strongly disfavoured \citep{Paoletti:2020ndu,PRL} (also by the studies on reionization sources \cite{2020ApJ...892..109N}). Moreover, both CMB data and their combination with astrophysical one show a preference for simple monotonic models of reionization disfavouring step-like and multiple burst models \citep{Paoletti:2020ndu, PRL}.

The lower   optical depth  from Planck 2018 data places reionization in a redshift range in substantial overlap with astrophysical  probes of reionization as UV luminosity density and IGM status from high redshift sources. 
This opens a scenario where the three sectors of data on reionization, UV, QHII and CMB can be combined to describe the reionization process from its onset to its end. The complementarity among the three different sectors is the strength of this treatment that is twofold. On one side it aims to tighten the constraints on the history of reionization but on the other hand the improved knowledge of reionization thanks to the astrophysical datasets removes many degeneracies \citep{Efstathiou:1998xx,Howlett:2012mh} and decreases the uncertainties on the cosmological model~\citep{PRL}. 

In this perspective is of crucial importance to account for the full cosmological model together with the history of reionization in a way to profile possible degeneracies and correlations among the parameters, previous analyses with fixed cosmological model include \cite{Robertson2010:Nature,Mitra2011,Mitra2015,Bouwens2015,Robertson2015,Ishigaki2015,Price2016,Gorce2017,Ishigaki2018,Mitra2018}. 
Our treatment using also CMB data allows jointly sampling over both the cosmological model and the reionization history, therefore both cosmological and reionization parameter constraints are guided by the data without a-priori priors.

In the present work we expand on our previous treatment\citep{PRL} increasing the degrees of freedom of the model and testing different data combinations. The paper is organized as follows. In section II we will present the treatment. In section III we will present the results for the different data and model extensions and in section IV we draw our conclusions.  

\section{Non-parametric reconstruction of reionization}
There are different approaches to treat reionization and its role in cosmological model constraints, the most commonly used are physical models as in  \cite{Planck2016:reion,Choudhury:2020kzh}
or a pure free form data fitting of the ionization fraction with a principal component analysis~\cite{Heinrich:2016ojb,Heinrich:2018btc}, to hybrid parametric fitting models~\cite{Paoletti:2020ndu,HS17,Millea:2018bko,Planck2018:param,Mortonson2007,Douspis2015}.
This paper focuses on the joint astrophysical and CMB data driven reconstruction based on the method developed in \citep{PRL} which we summarize below (for alternative  data driven reconstructions see also \cite{Gorce2017,Mitra2018,Mitra:2011uv,Qin:2020xrg}).
We focus on averaged reionization leaving the patchy case for a future work (for patchy reionization effects on CMB see for example \citep{Smith:2016lnt,Roy:2018gcv}).

\subsection{General framework}
The challenge of combining astrophysical data and the CMB is mainly due to a compatibility issue with the UV luminosity density data. 
If both CMB and QHII data can be combined within the same methodology, namely a model of the ionization fraction function holds for both the datasets,
UV luminosity density data are of a different nature. They trace the evolution of the sources of the reionization not the ionization fraction itself. 
It is necessary to change  methodology in the treatment of the reionization history and this change of paradigm is the core of this kind of treatment: 
reionization is no longer described by a parametric model of the ionization fraction with time but it is solved from its basic equation. 
The new variables at play become the sources and sinks namely the ionizing UV luminosity density and the recombination time. 
These quantities enter in the volume filling factor equation:
\beq
\frac{dQ_{\rm HII}}{dt}=\frac{\dot{n}_{\rm ion}}{\langle n_{\rm H}\rangle}-\frac{Q_{\rm HII}}{t_{\rm rec}},\,
\eeq
which we solve for our reionization history.

The source term is represented by the ionizing photon production rate,~$\dot{n}_{\rm ion}$, that is the product of the UV luminosity density  $\rho_{\rm UV}$, which represents a tracer of the star formation, the efficiency of conversion of UV luminosity into ionizing photons $\xi_{\rm ion}$ and the escape fraction $f_{\rm esc}$ that represents the fraction of ionizing photons that actually reach the IGM and proceed with the reionization.
Current data do not provide sufficient information on the latter parameters to have grip in constraining them, for this reason in our previous work~\cite{PRL} we fixed their value to $\log_{10}\langle ( f_{\rm esc}\xi_{\rm ion}\rangle/[{\rm erg^{-1} Hz}] ) =24.85$, consistent with other analyses~\cite{Ishigaki2015,Price2016,Ishigaki2018,Madau2017}. 
The assumptions for the description of the source term will be one of the main focus of our analysis, in fact, we will test both the fixed efficiency assumption and we will test a different assumption in the derivation of the UV luminosity density from the measured UV luminosity function.

The recombination time is the other variable of our framework and is defined by $t_{\rm rec}=1/\l[C_{\rm HII}\alpha_{\rm B}(T) (1+Y_p/(4X_p))\langle n_{\rm H}\rangle (1+z)^3\r]$
with $C_{\rm HII}$ as the clumping factor, $\alpha_{\rm B}(T)$ the recombination coefficient, $\langle n_{\rm H}\rangle$ the  density of hydrogen  and $X_p$, $Y_p$ as the hydrogen and helium abundances.

\subsection{Operational configuration}

The reconstruction is performed within a modular structure that partitions the redshift range in different bins in order to consider also step like and non monotonic reionization histories.  
Each bin comprises of three different variables: the bin position in redshift $z_{\rm int}$, whose range is conditional to the number of bins assumed, the UV luminosity density $\rho_{\rm UV}$ and the recombination time $t_{\rm rec}$.
The UV luminosity density and the recombination time are interpolated between different redshifts through Piecewise Cubic Hermite Interpolating Polynomials (PCHIP) that guarantee the required freedom for the variable quantities to be guided by the data. 
The node structure is incremental in complexity of the possible reionization histories increasing the freedom of the reconstruction. Since we demonstrated that Bayesian evidence disfavours reionization histories beyond a simple monotonic single step~\cite{PRL} we focus in this work on single bin analyses with only some tests on the three bins case. We assume best fit double power law function for the luminosity density beyond redshift ranges covered by the priors on $z_{\rm int}$. Note that double power law function has recently been shown to be in agreement with the data also with Gaussian Process reconstruction~\cite{Krishak:2021fxp}. Note that this does not reduce the flexibility of our model as the values of $\rho_{\rm UV}$ at different times denoted by $z_{\rm int}$ are allowed to vary freely in a wide uniform prior range.

The non-monotonic histories are represented by the three nodes case named $B3$. Monotonic histories are represented by the single bin treatment named $B1$, that allows also for step like histories, and the minimal case with a single node with the constraint $\dot{n}_{\rm ion}t_{\rm rec}=\langle n_{\rm H}\rangle$ at $z=a_{\rm int}$ that restricts reionization to a single burst. For Helium reionization we assume the default hyperbolic tangent provided by the Boltzmann code~\citep{Lewis1999} (note however that the study of Helium reionization is in rapid development \cite{Lau:2020chu,Linder:2020aru,Caleb:2019apf}).

We assume also some boundary conditions on the onset and the end of reionization in order to ensure the physicality of the reconstructed histories. In particular, we ensure that reionization is over by $z=5.5$, conservatively allowing for still a residual contribution from $z=6$~\citep{Schroeder2013}, and it remains over until $z=0$. As highest redshift for the onset of reionization we assume $z=30$ which leaves room for early reionization, although disfavoured by data.
The boundary nodes are configured to be consistent to best fit logarithmic double power law (see, Eq. (39) of~\cite{Ishigaki2015}) and also with~\cite{Becker2013}.

The usual parametrization for the CMB, the integrated optical depth, in this framework is a derived parameter computed from the volume filling factor integrated from the beginning of reionization:
\begin{equation}
\tau=\int_{0}^{z_{\rm begin}}\frac{c(1+z)^2}{H(z)}Q_{\rm HII}(z)\sigma_{\rm Thomson} \langle n_{\rm H}\rangle\Big(1+\frac{Y_p}{4X_p}\Big)
\end{equation}
with $\sigma_{\rm Thomson}$ is the Thomson scattering cross-section. 

The parameters of the cosmological model are jointly varied with the reionization history, therefore accounting for possible correlations between them~\cite{HPFS18,Paoletti:2020ndu}. We therefore vary the five parameters of the standard model, the optical depth being already included as derived parameter in the reionization reconstruction: the baryon density $\Omega_{\rm b}h^2$, the dark matter density $\Omega_{\rm CDM}h^2$, the angular diameter distance to the horizon at recombination $\theta$, the amplitude of primordial fluctuations $A_{\rm s}$, the tilt of the primordial fluctuations power spectrum $n_{\rm s}$. To explore the parameter space and compute the Bayesian posterior distributions of the parameters we use the Markov Chain MonteCarlo sampler {\tt CosmoMC}~\cite{Lewis2002}, for the CMB data we use the Planck 2018 likelihood suite~\cite{Planck2018:like} which requires the co-sampling of some nuisance foregrounds and calibration parameters. 

\subsection{Data and priors}

We now detail the baseline data combination we used, which are already used in~\citep{PRL}.

Concerning the UV luminosity density the data are derived from the observations of six galaxy clusters in the  Hubble Frontier Field~\cite{HFFsite,Lotz2017} going from redshift of 6 to 11~\cite{Bouwens2014,Ishigaki2018}. To extract the density we assume a Schechter model and integrate the UV luminosity function up to a magnitude of $-17$. This is a conservative choice, in fact, it cuts out the low luminosity tail which, being populated by low luminosity sources, may represent a non negligible contribution but at the same time the data which we use in this work seem to indicate a change in the shape in this region, with a possible acceleration still to be confirmed by future data (see figure 8 of \cite{Ishigaki2018}).

In our baseline data combination we consider also the  measurements of the ionization state of the IGM around high redshift sources. 
We use the available data from Quasars and Gamma Ray Burst in a redshift range in substantial overlap with the UV luminosity density data, from $z=6$ to $z=8$. The complete dataset that we use here can be found in the following references:~\cite{McGreer:2014qwa,Totani:2005ng,McQuinn:2007gm,Schroeder2013,Greig:2016vpu,Davies:2018pdw,McQuinn:2007dy,Ouchi2010,Schenker2014,Tilvi2014,Mason:2017eqr,Mason:2019ixe}.
\begin{figure*}
\includegraphics[width=0.3\textwidth]{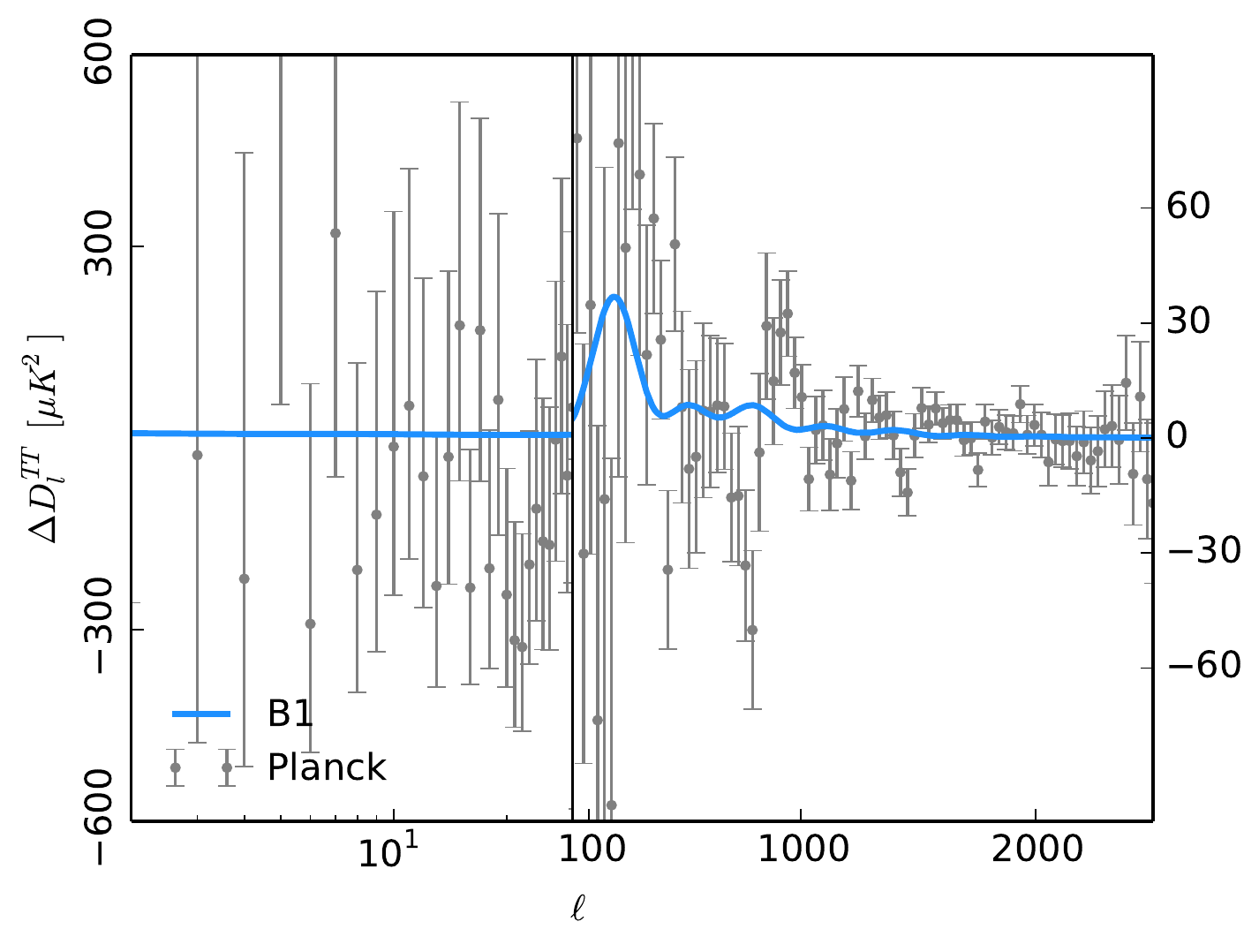}
\includegraphics[width=0.3\textwidth]{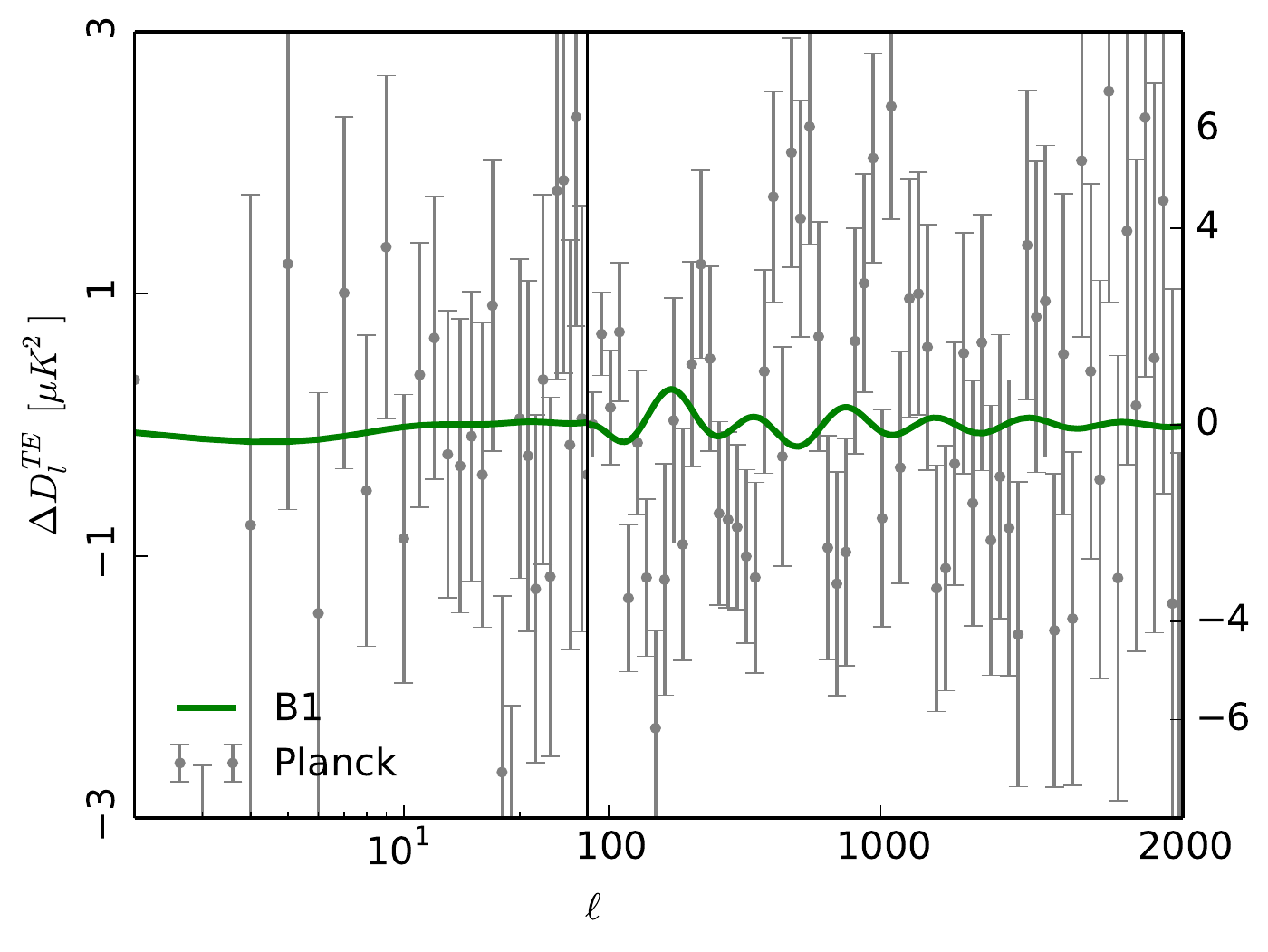}
\includegraphics[width=0.3\textwidth]{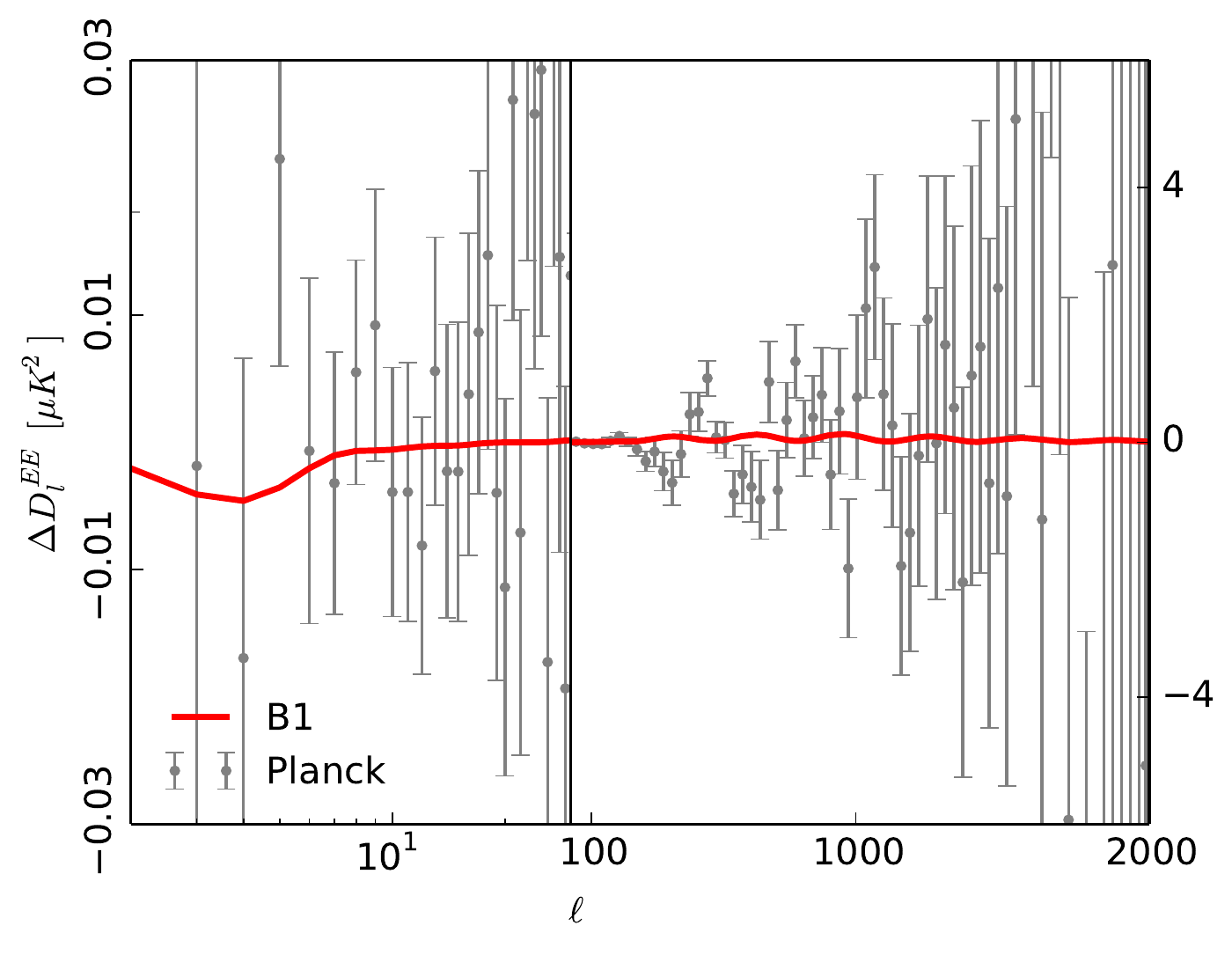}
\caption{Residuals of the B1 case best fit, following our previous set up \citep{PRL}, from the full data combination with respect to the Planck best fit, with the hyperbolic tangent model, compared with the data points from Planck. Left panel is TT, middle panel is TE and right panel is EE.}
\label{Spect}
\end{figure*}

For the CMB we use the Planck 2018 data baseline which includes temperature, polarization and lensing~\citep{Akrami:2018vks,Planck2018:like}.
For the temperature anisotropies the likelihood on large multipoles is based on auto and cross spectra of the high frequency instrument, whereas for the low multipoles it is uses a Gibbs sampling based on the component separated {\texttt{commander}} map. The Planck 2018 temperature data are mostly incremental on the 2015 release with the main changes on large angular scales where no ancillary data but only Planck is used for the component separated map. Polarization instead has improved on both large and small angular scales. A better characterization of the temperature to polarization leakage guarantees a cosmology-grade high-$\ell$ polarization in the 2018 release. The large angular scales are now based on the 100x143 GHz cross spectrum with a simulation based likelihood which considers only the E-mode polarization and not the cross-correlation with temperature. The lower noise level of the cross spectrum has halved the uncertainty on  the optical depth in the standard hyperbolic tangent scenario~\cite{Planck2018:param}. Together with primary anisotropies the Planck baseline includes the weak lensing of the CMB in the conservative range $8 \le \ell \le 400$~\citep{Planck2018:lensing}.

Concerning the priors for the redshift ranges of the bins in the B3 case the conditional priors are chosen to maximize the relevance of the different data: the first node is in the range $z=5.5-8$ covering the QHII data, the second in $z=8-12$ for tracing possible features in the luminosity function of the UV and the third node spans the high redshift range $z=12-30$ to be sensitive to the CMB contribution. The cases with a single node its position spans the entire range $z=5.5-30$.

In~\autoref{Spect} are displayed the changes induced by the resulting reconstruction of the reionization for the B1 from the full data combination on the CMB anisotropy angular power spectra in temperature and polarization compared with the current measurements by Planck 2018. The residuals with respect to the Planck 2018 best fit, based on the hyperbolic tangent model, show an impact on both temperature on the intermediate scales and on large angular scales in polarization. In the middle panel we display also the temperature and polarization cross-correlation.

\section{Results}

In this section we present the results of the analyses performed by using both different model assumptions and different data.
\subsection{Changing the model}

\subsubsection{Single node-B0: $\Lambda CDM$+$A_{L}$}

We consider now a simple extension on our previous work~\cite{PRL}, by using the same assumptions we want to verify if a reconstructed reionization history which is based on the combination of astrophysical data has an impact on one of the main {\it curiosities} of current Planck 2018 results:the higher value of the lensing amplitude from the CMB angular power spectrum.
In this model the effect of the lensing on the CMB anisotropies angular power spectrum is mediated by an overall amplitude parameter $A_{L}$ which does not represent a physical parameter but a simple overall rescaling. This extension is a good tracer of possible residuals systematics in the small angular scale region and due to its effect on the angular power spectrum may be strongly degenerate with the model of reionization. Therefore, we investigate if a different reionization dynamics with respect to the hyperbolic tangent used in the Planck baseline results may lead to a different value of the lensing amplitude.
Since the one bin settings B0 and B1 are equivalent for this specific case we assume the B0 for this analysis. 
The resulting constraints for both the cases of including or not the lensing likelihood, based on the extraction of the physical lensing signal from maps, are shown in~\autoref{Table2}. The result for the case without the lensing likelihood is $A_{L}=1.176_{-0.063}^{+0.064}$ and we do not note a significant impact of the reionization reconstruction with respect to the standard hyperbolic tangent case that provides $A_{L}=1.180\pm 0.065$~\citep{Planck2018:param}, there is just a fraction of $\sigma$ reduction of the central value. The case which includes the physical lensing likelihood gets back the central value to unity as in the standard hyperbolic tangent case. We can therefore conclude that the lensing amplitude model is mostly sensitive to the integrated optical depth and not to the history of reionization. This is coherent with a main effect from the lensing on the small angular scales in the acoustic peak region where the reionization effect is mainly mediated by the optical depth whereas the reionization history has a greater effect on the large angular scales in E-mode polarization.
\begin{table}
\begin{tabular}{ |p{2.3cm}||p{3cm}|p{3cm}|  }
 \hline
 \multicolumn{3}{|c|}{B0} \\
 \hline
Parameters & Planck 2018-no lensing +UV17+QHII & Planck 2018 +UV17+QHII\\ \hline
$\Omega_b h^2$ & $0.0226\pm{0.0002}$& $0.0225\pm{0.0002}$\\
$\Omega_c h^2$ & $0.118\pm{0.002}$& $0.118\pm{0.001}$\\
$100\theta_{MC}$ & $1.0411\pm{0.0003}$& $1.0411\pm{0.0003}$\\
$\tau$ & $0.0515_{-0.0012}^{+0.0010}$&$0.0512_{-0.0012}^{+0.0010}$\\
$A_{L}$ & $1.176_{-0.063}^{+0.064}$& $1.067\pm{0.036}$\\
$z_{int}^1$ & $21.5_{-3.2}^{+6.7}$& $21.6_{-3.1}^{+6.8}$\\
\makecell[l]{$\log_{10}[\rho_{\rm UV}]$\\\tiny{$[{\rm erg}/({\rm s}\, {\rm H_z} {\rm Mpc}^{3})]$}} & $21.1_{-2.2}^{+1.9}$& $21.1_{-2.4}^{+1.8}$\\
${\rm{ln}}(10^{10} A_s)$  & $3.033\pm{0.0006}$& $3.033\pm{0.0006}$\\
$n_s$ & $0.9709_{-0.0048}^{+0.0048}$& $0.9693_{-0.0046}^{+0.0047}$\\
 \hline
\end{tabular}
\caption{~\label{Table2}Constraints on the cosmological parameters for the $A_{lens}$ extension using a B0 configuration. Error bars are the 68\% C.L..}
\end{table}

\subsubsection{Relaxing the assumptions on the source term}

Current astrophysical data do not provide direct information on the escape fraction, and the CMB is totally insensitive to it. On this basis in our previous work~\citep{PRL} we assumed a fixed value for the product of photon conversion efficiency and the escape fraction, $\log_{10}\langle f_{\rm esc}\xi_{\rm ion}\rangle/[{\rm erg^{-1} H_z}]=24.85$. A preliminary test in \cite{PRL} did not show a significant impact on the analysis but we will now extensively test this result.
In the following results of the paper we will always lift the assumption of fixed $f_{\rm esc}\xi_{\rm ion}$ and let the product of the conversion efficiency and the escape fraction free to vary.\\

We now test extensively the impact of the removal of this assumptions on the reionization reconstruction using the baseline data combinations previously described.
With this additional degree of freedom we do not consider the CMB only data option because it would leave the efficiency totally unconstrained and would not provide any additional information up to the point of a difficult convergence of the MCMC.\\

\begin{figure}
\includegraphics[width=0.5\textwidth]{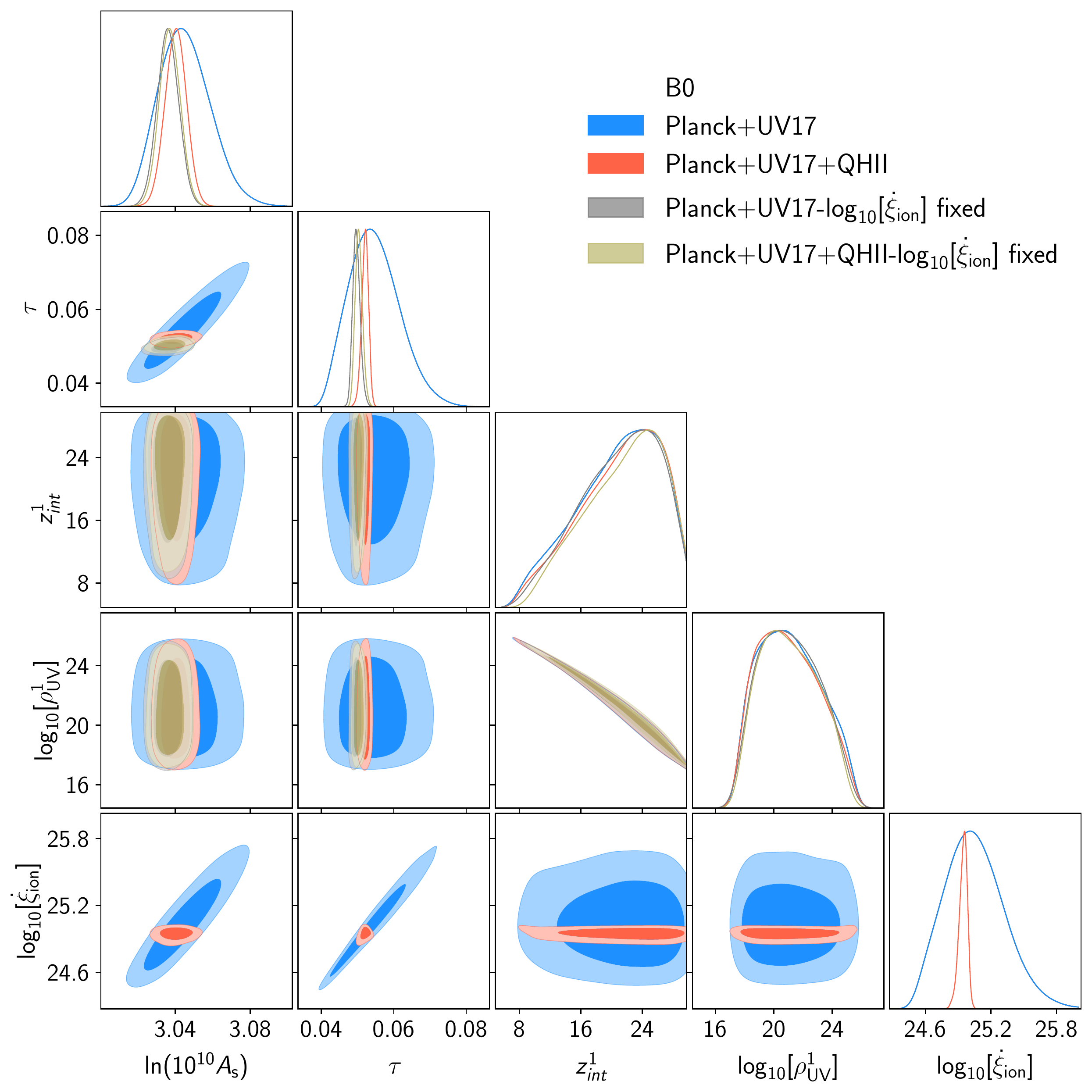}
\caption{~\label{F2}Constraints on reionization parameters for the B0 case. We compare the case varying the efficiency for the combination of CMB and UV luminosity density data and the addition of QHII data with the same cases with fixed efficiency. }
\end{figure}

We start with a preliminary case by analysing the simplest configuration, the B0, namely a single burst monotonic reionization.
In~\autoref{F2} we present the two dimensional marginalized posterior distributions for the reionization parameters. We compare the results of the CMB+UV and CMB+UV+QHII data combinations (in blue and red) with the corresponding cases where $\log_{10}\langle f_{\rm esc}\xi_{\rm ion}\rangle/[{\rm erg^{-1} H_z}]=\log_{10}[\dot{\xi}_{\rm ion}]/[\rm{ erg^{-1} H_z}]=24.85$~\cite{PRL} (in gray and khaki colors). The constraints on all cosmological parameters are in~\autoref{Base}. The standard cosmological parameters are stable with respect to the reconstruction of reionization, when compared with the standard hyperbolic tangent case, with only a marginal shift in the scalar spectral index of a fraction of sigma. The results for the reionization parameters are pretty stable compared with the fixed escape fraction case with an increase of the central value of the optical depth which goes to $\tau=0.0547_{-0.0078}^{+0.0060}$ and $\tau=0.0521_{-0.0008}^{+0.0010}$ to be compared with the results with fixed efficiency which provides $\tau=0.050\pm 0.001$ and $\tau=0.051\pm 0.001$.
In both data combinations there is a sensible increase of the error bars for the integrated optical depth especially for the CMB+UV case. The addition of QHII data breaks the degeneracies providing also a good constraint on the escape fraction $\log_{10}[\dot{\xi}_{\rm ion}]/[{\rm erg^{-1} H_z}]=24.95_{-0.03\,-0.08}^{+0.04\,+0.07}$ 
which is compatible at three sigmas with the value we used in previous analyses.\\
\begin{table}
\begin{tabular}{ |p{2.3cm}||p{2.5cm}|p{2.7cm}|  }
 \hline
 \multicolumn{3}{|c|}{B0} \\
 \hline
Parameters &  Planck 2018+UV17& Planck 2018+UV17+QHII\\
 \hline
$\Omega_b h^2$ & $0.0224\pm 0.0001$& $0.0224\pm 0.0001$\\
$\Omega_c h^2$ & $0.120\pm 0.001$& $0.120\pm 0.001$\\
$100\theta_{MC}$ & $1.0409\pm 0.0003$& $1.0409\pm 0.0003$\\
$\tau$ & $0.0547_{-0.0078}^{+0.0060}$&$0.0521_{-0.0008}^{+0.0010}$\\
$z_{int}^1$ & $20.8_{-3.4}^{+7.3}$& $21.0_{-3.3}^{+7.2}$\\
\makecell[l]{$\log_{10}[\rho_{\rm UV}]$\\\tiny{$[{\rm erg}/({\rm s}\, {\rm H_z} {\rm Mpc}^{3})]$}}& $21.1_{-2.8}^{+1.7}$& $21.0_{-2.7}^{+1.7}$\\
\makecell[l]{$\log_{10}[\dot{\xi}_{\rm ion}]$\\\tiny{$[{\rm erg\, H_z^{-1}}]$}} & $25.05_{-0.30}^{+0.24}$& $24.95_{-0.03}^{+0.04}$\\
${\rm{ln}}(10^{10} A_s)$ &$3.045_{-0.015}^{+0.012}$& $3.040_{-0.006}^{+0.006}$\\
$n_s$ & $0.9650\pm 0.0040$& $0.9644\pm 0.0039$\\
$ \Delta_z^{\rm Reion}$ & $2.78_{-0.15}^{+0.12}$& $2.80_{-0.15}^{+0.11}$\\
 \hline
\end{tabular}
\caption{~\label{Base}Constraints on the parameters for the single burst reionization case B0 when the efficiency is let free to vary. The error bars are the 68\% C.L.}
\end{table}

\begin{figure}
\includegraphics[width=0.5\textwidth]{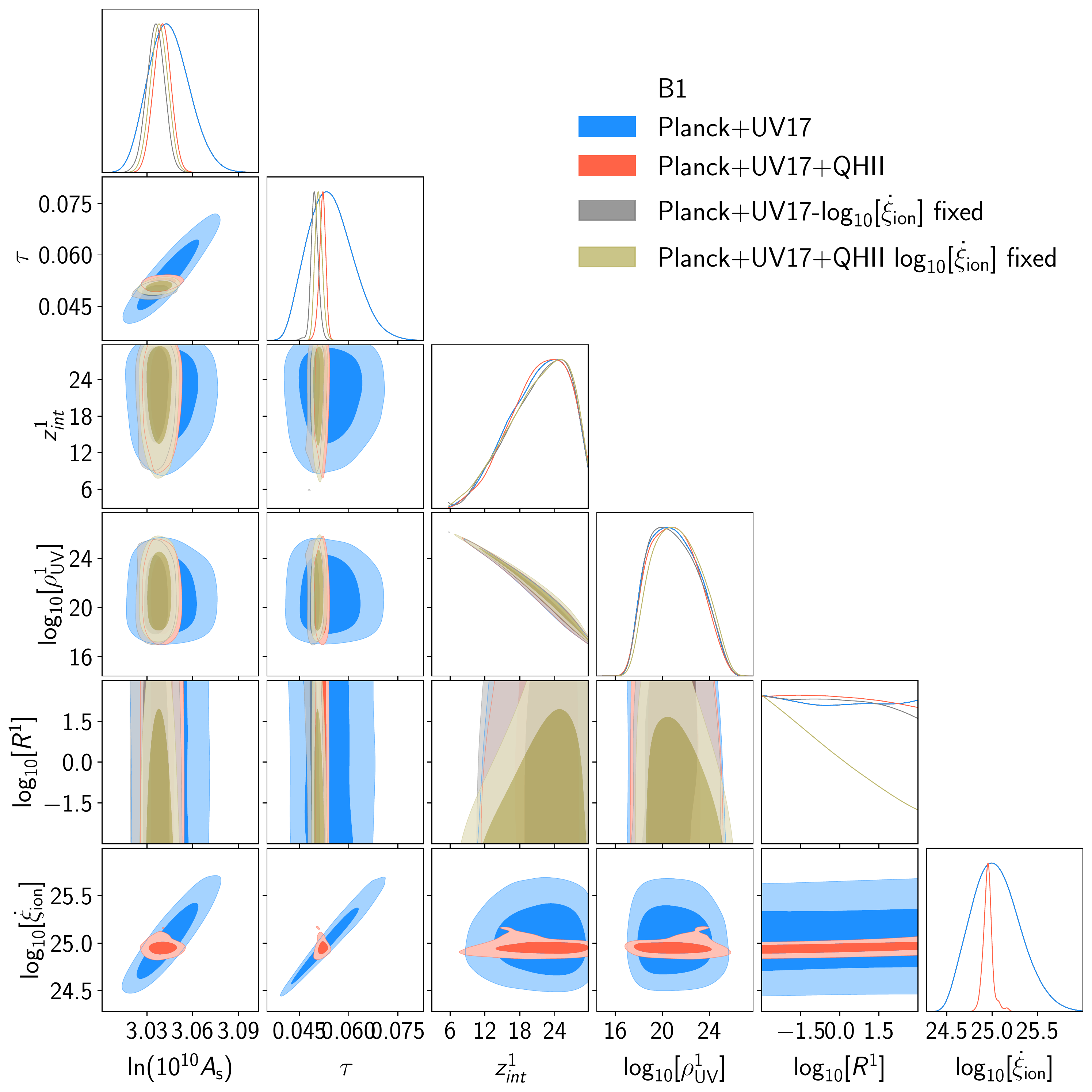}\caption{~\label{F4}Constraints on reionization parameters for the B1 case. We compare the case varying the efficiency for the combination of CMB and UV luminosity density data and the addition of QHII data with the same cases with fixed efficiency.}
\end{figure}

We go into the details of the difference between fixed and free efficiency model increasing the complexity to the B1 case where we allow for possible steps in the reionization process leaving free to vary also the recombination time.
The two dimensional posteriors for the reionization parameters are shown in~\autoref{F4},using the same color scheme of B0, kakhi and gray for the fixed efficiency and red and blue for the free case, and the constraints on the cosmological and reionization parameters are in~\autoref{Table3}. We note a similar pattern to the single burst case. 
The additional freedom of the escape fraction only marginally impacts the constraints on cosmological and reionization parameters and in the full data combination we constrain the escape fraction in agreement at 95\% C.L. with our previous assumption 
$\log_{10}[\dot{\xi}_{\rm ion}]/[{\rm erg^{-1} H_z}]=24.95_{-0.05 \,-0.11}^{+0.04 \,+0.12}$.
As for the B0 also for the B1 we note a significant higher optical depth with respect to the fixed efficiency case, especially in the CMB+UV case but the larger error bars make it compatible with our previous results within 68\% C.L.. We also note that the additional degree of freedom of the recombination time does not affect the uncertainties in the parameters being the recombination time unconstrained.

\begin{figure}
\includegraphics[width=0.5\textwidth]{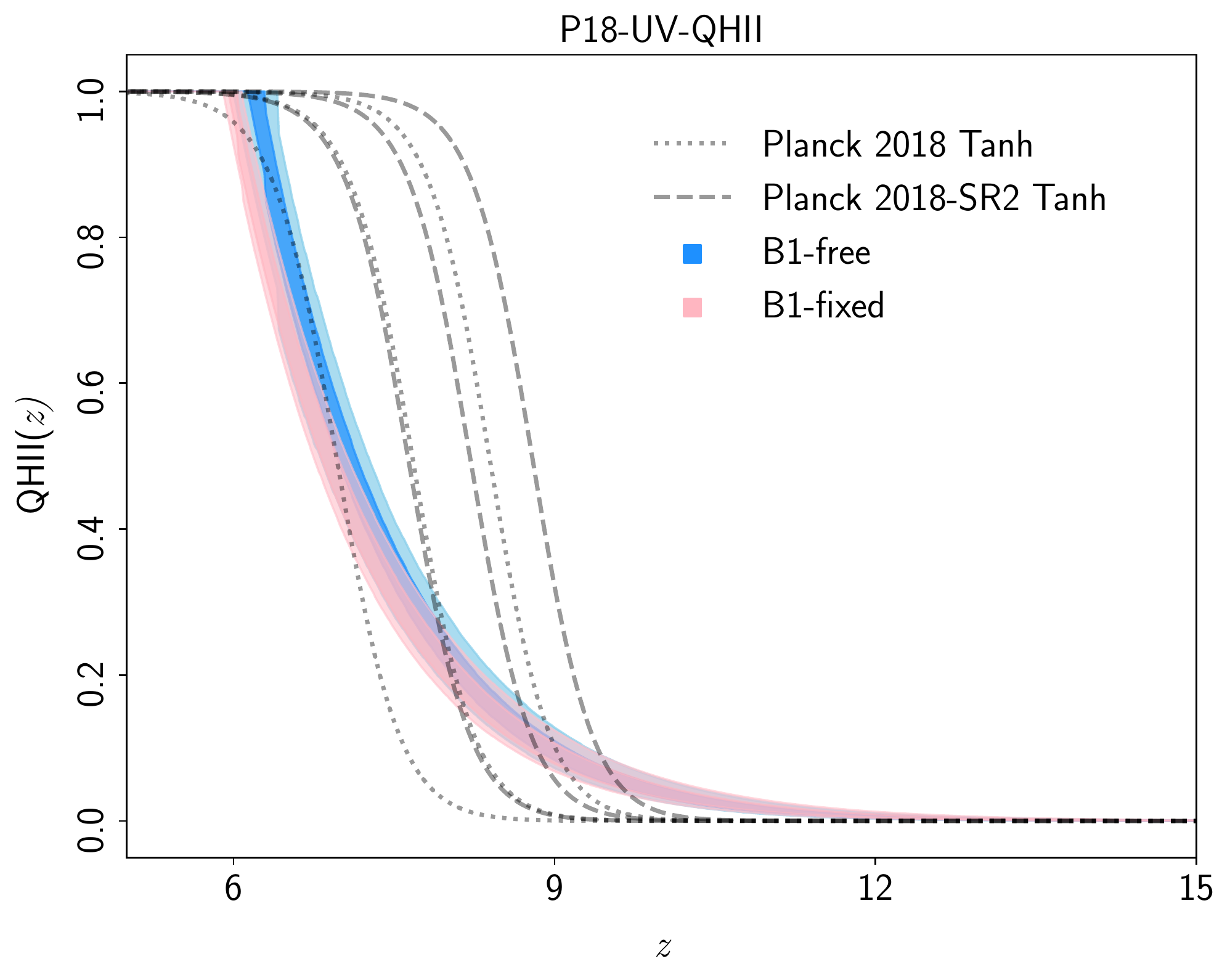}
\caption{~\label{Rec21}Reconstructed reionization history using the full data combination of astrophysical data and CMB. We show the comparison for the B1 case between fixed and varying $\dot{\xi}_{\rm ion}$.}
\end{figure}

In ~\autoref{Rec21} we present the comparison of the reconstructed reionization histories using the full combination of astrophysical and CMB data between the case of free and fixed $\dot{\xi}_{\rm ion}$. We note a significant overlap between the two histories especially in the higher redsfhit region where the free efficiency does not impact the reconstruction. On the other hand in the second part of the reionization the free efficiency prefers a steeper rise which brings to an earlier completion of reionization with respect to the case where we assumed a fixed value. This is reflected also in the slightly higher value of the reconstructed optical depth.
Considering the CMB alone case is not capable to provide constraints due to the additional freedom of the efficiency we now investigate the relevance of the current CMB data beyond their importance for the cosmological model focusing specifically on their impact on the reconstruction of reionization. To this purpose we analysed the combination of both astrophysical datasets with some of different subsets of Planck 2018 data.

\begin{figure}
\includegraphics[width=0.5\textwidth]{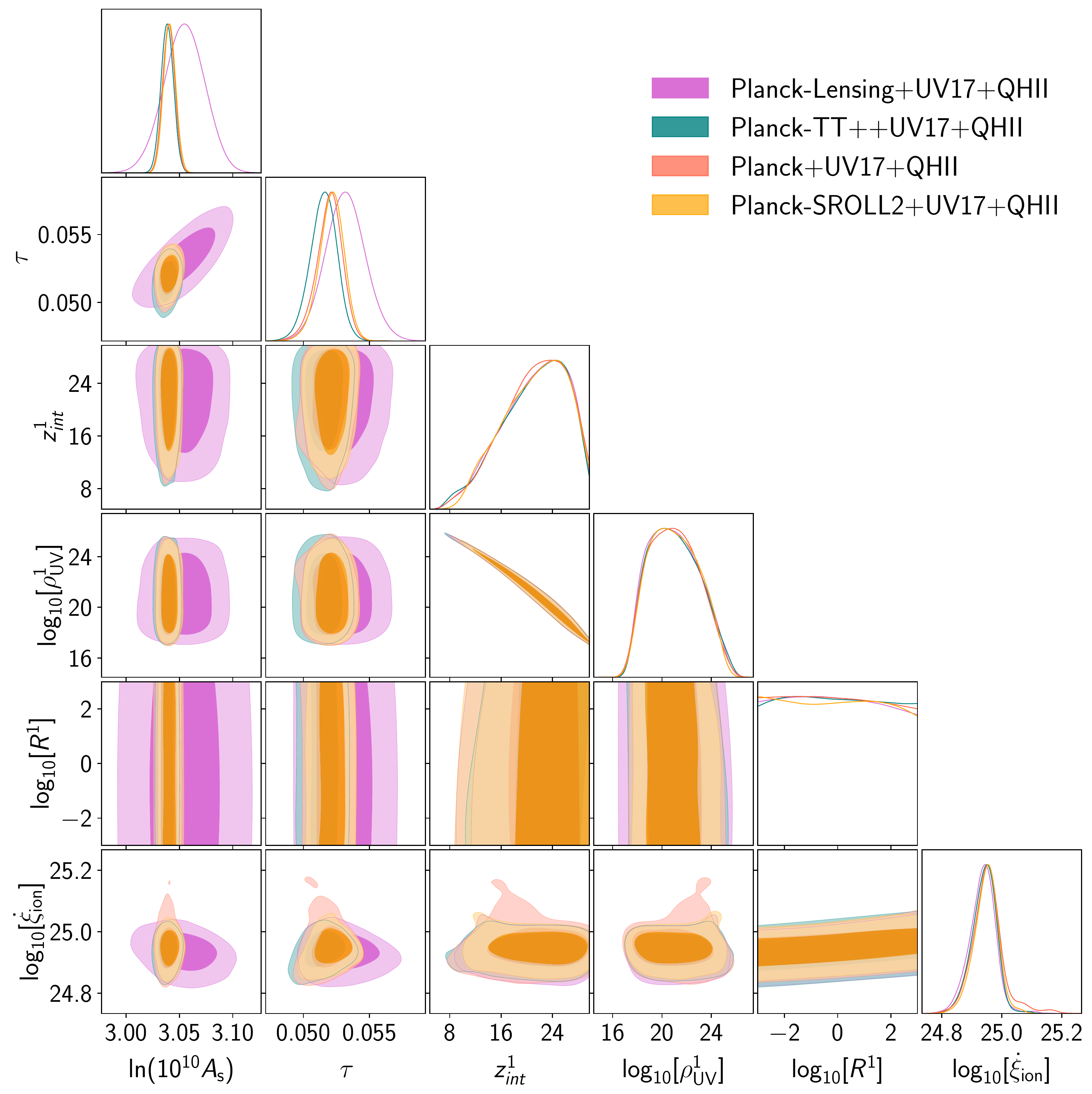}
\caption{~\label{FC1} Two dimensional contours for the reionization parameters, we compare the combination of UV luminosity density and QHII with different Planck 2018 sub-datasets.}
\end{figure}

In the fourth column of~\autoref{Table3} and in orange contours in~\autoref{FC1}  we show additional results obtained by switching the low-$\ell$ polarization likelihood of Planck 2018 with the more recent SROLL2 version \citep{Delouis:2019bub,Pagano:2019tci}, based on a map-making algorithm that provides a reduced level of systematics. In the standard hyperbolic tangent case the use of SROLL2 likelihood implies an higher optical depth of a bit less than a sigma ($\tau = 0.059 \pm 0.006$ with respect to the Planck 2018 baseline $\tau=0.054\pm 0.007$), also for the reconstruction we a have slight increase of the optical depth but only of a fraction of sigma. In a framework like the reconstruction with the combination of CMB and astrophysical data the history of reionization is less sensitive to the CMB with respect to the hyperbolic tangent which instead being parametrized only by the optical depth is more sensitive to changes in the CMB EE-power spectrum. 

We also considered the case excluding the Planck polarization data both at large and small angular scales, shown in teal contours in ~\autoref{FC1} and the case with only the lensing likelihood \footnote{Due to the reduced constraining power for this specific case we fixed the scalar spectral index, the angular diameter distance to the last scattering and the baryon density to the Planck 2018 baseline values} shown in orchid contours. The parameter constraints are presented in the last two columns of~\autoref{Table3}.
We remark how also for these data subsets without contribution of the E-mode polarization we have a very good recovery of the baseline case. The temperature alone results provides an optical depth in agreement with the baseline case, as expected considering that in this case the main driver for the reionization reconstruction are the astrophysical data. For the lensing only case the results show a slightly larger reconstructed optical depth but provides similar constraints for the reionization parameters. We remind anyway that this case is affected by the fixing of good part of the cosmological parameters to their Planck baseline value which may bias the results towards that case. This analysis show how the main driver of the reionization reconstruction are the astrophysical data when the efficiency is left free to vary. Anyway we stress that reionization is just one of the pawns in the game when constraining cosmological models and that the CMB remains a fundamental probe to constrain the unique entity composed by the cosmological and reionization model. Moreover future data with a cosmic variance limited sensitivity in the E-mode polarization will probably change the current scenario but we leave this forecasts for a future work.

\begin{figure}
\includegraphics[width=0.5\textwidth]{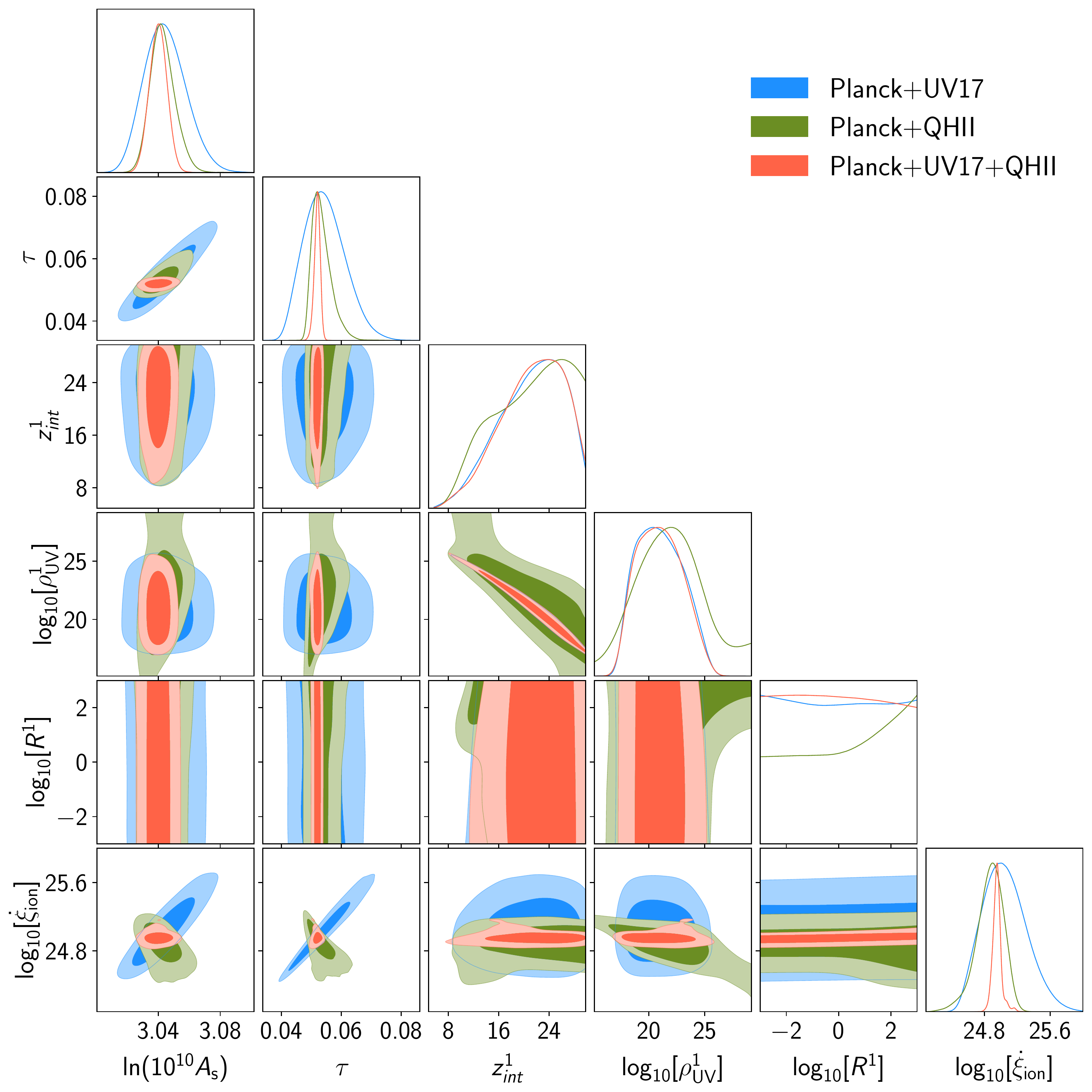}
\caption{~\label{FC2} Two dimensional contours for the reionization parameters, we compare the three combinations of CMB with astrophysical datasets.}
\end{figure}

\begin{figure*}
\includegraphics[width=\textwidth]{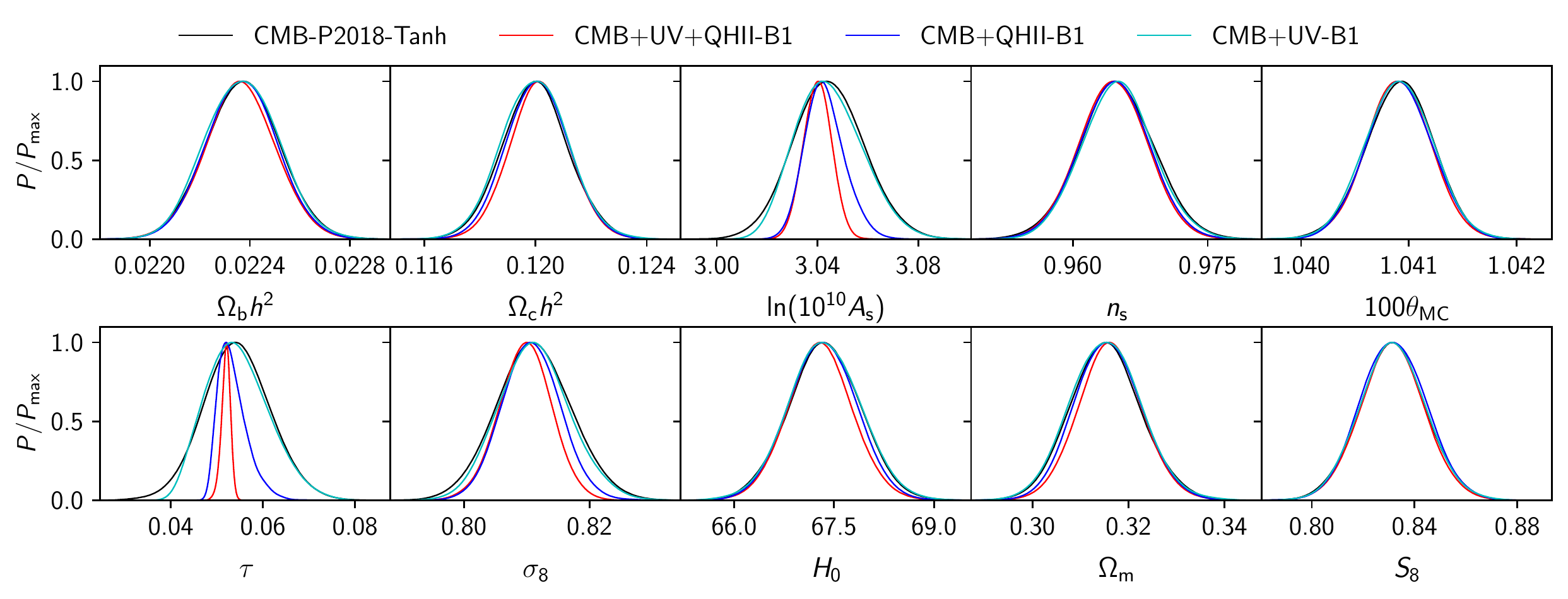}
\caption{~\label{1d} Cosmological parameters constraints comparison among the different combinations of datasets and with respect to Planck 2018-baseline hyperbolic tangent results.}
\end{figure*}

\begin{figure*}
\includegraphics[width=\textwidth]{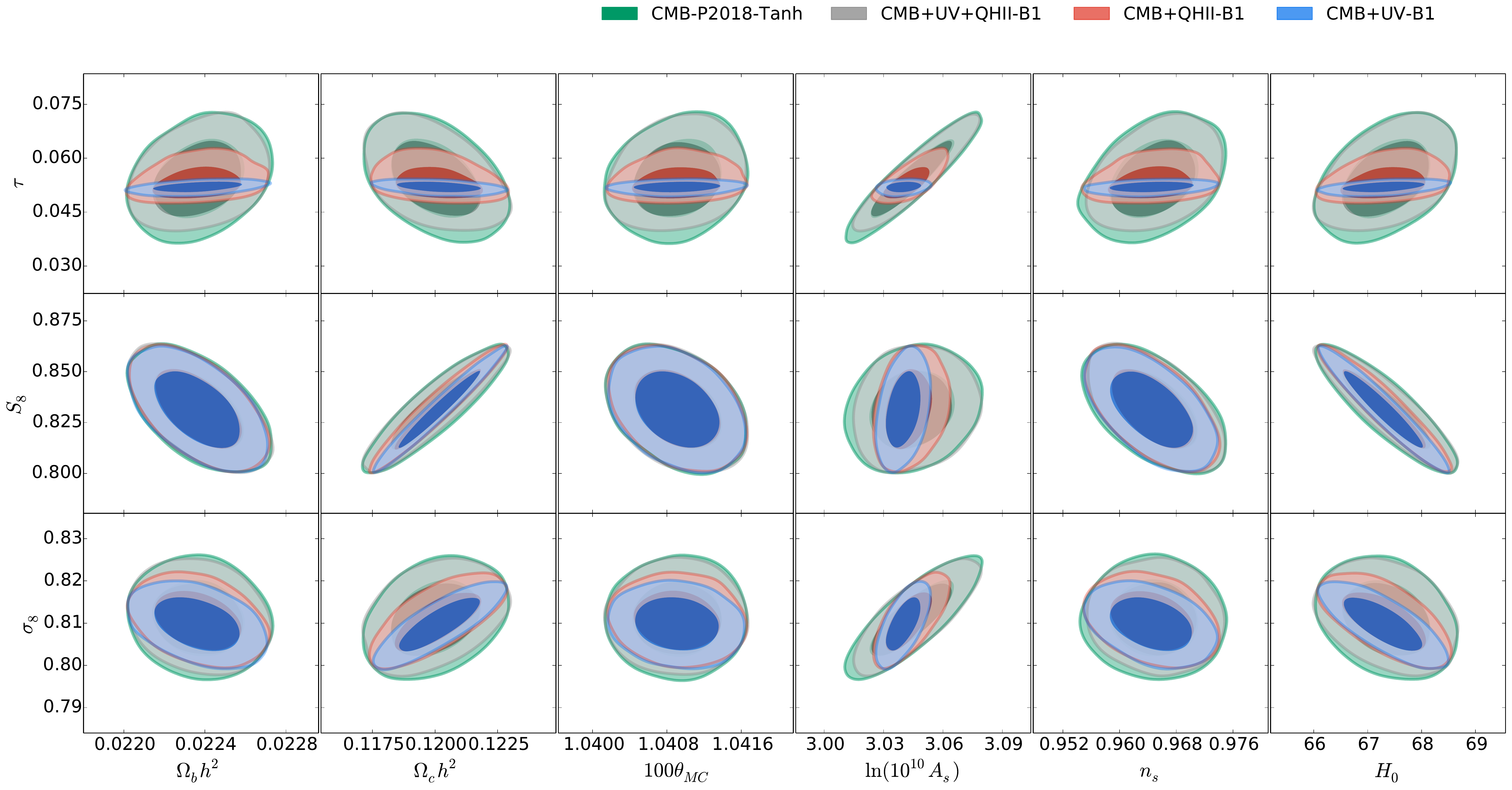}
\caption{~\label{2d} Cosmological parameters two-dimensional posterior distributions for the different combinations of datasets compared with Planck 2018-baseline hyperbolic tangent results.}
\end{figure*}

In order to investigate the two different pulls of the two astrophysical data in combination with the CMB we compare the full combination with the combinations of CMB with either UV luminosity density or QHII. The results are presented in~\autoref{FC2} and we first note the agreement of the separate astrophysical data combinations with CMB that do not show any sensible bias within the enlarged error bars. The QHII data provide tighter constraints on the optical depth with a slightly lower central value. The addition of UV luminosity density data further tighten the constraints which remain in agreement with the single data combination. We note the change in the degeneracy direction for  the efficiency and the optical depth between the CMB+UV and CMB+QHII combinations. 
This change is mainly due the addition of QHII fixing the ionization fraction at fixed redshift which causes the optical depth to decrease with increasing efficiency to maintain these fixed points.
In \autoref{1d} and \autoref{2d} we show the one and two dimensional posterior distributions for the other cosmological parameters compared among the three data combination and with the standard Planck results with the hyperbolic tangent.
We do not note any significant shift in the cosmological parameters, the only effect is the reduction of the error bars due to the addition of astrophysical dataset especially for the parameters which have larger degeneracies with the optical depth as the amplitude of primordial fluctuations and the $\sigma_8$.This follows on the line that this kind of astrophysical data reconstruction of reionization is decoupled from the standard cosmological model, at least for current data sensitivities.\\

\begin{figure}
\includegraphics[width=0.5\textwidth]{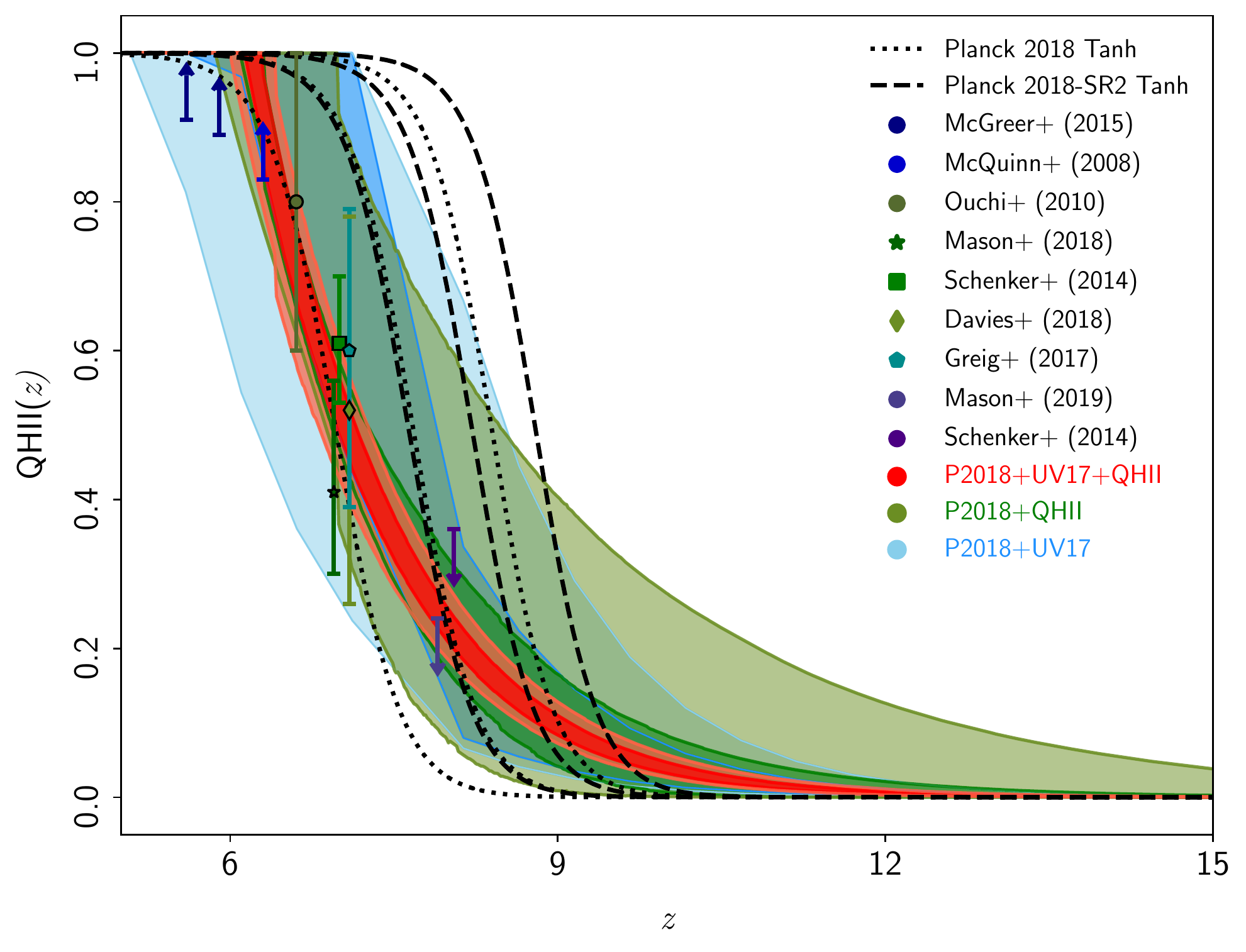}
\caption{~\label{B1A} Reconstructed reionization history bands at 68\% and 95\% C.L. for the three dataset combinations with CMB. Dashed and dotted lines represent the 68\% C.L. for the hyperbolic tangent of Planck 2018 baseline and SROLL2 results. Data points are QHII data used. Note that the sharp edges in the band for P2018+UV17 are just an artefact of the post processing.}
\end{figure}

In \autoref{B1A} we present the reconstructed reionization histories for the three data combination. The bands represent the 68\% and 95\% confidence levels. 
The reconstructed reionization histories are in agreement at 95\% C.L. although different data combinations provide slightly different scenarios.
This difference is mainly due to the pulls of the astrophysical data. The ionization fraction data from AGN and GRB strongly constrain the redshift range below 8 imposing a shallow slope in the rising towards $x_e=1$ of the end of reionization. This shallower rise is in agreement within error bars with the standard Planck 2018 baseline hyperbolic tangent but becomes more distant especially near the end of reionization when the SROLL2 data are used, due to the preference of SROLL2 for higher optical depth in the hyperbolic tangent model. This similarities in the reionization histories although providing different values of the optical depth show how the optical depth may be a good summary statistics of reionization for the purpose of constraining the cosmological model but fails to be a good representative of the reionization history. The single addition of the QHII data to the CMB leaves room for a longer tail at high redshift which is instead reduced when UV luminosity density are taken into account. The higher redshift data from the UV luminosity density reduce the allowed range for the onset of reionization but provide less stringent constraints in the intermediate-low redshift range that with larger error bars is compatible with the hyperbolic tangent for both cases of Planck 2018 data. The combination of these two different pulls on the reionization history produces the tight red contours which show a preference for a shallower reionization history but at the same time disfavour early onsets and are only marginally in agreement with the Planck 2018 baseline hyperbolic tangent and in slight tension with the SROLL2 case. Considering all the three combinations we can conclude that the main driver of the difference between the reconstructed reionization history and the standard hyperbolic tangent can be traced back to the QHII data pulling for a shallower rise in the $z<8$ region. With the expected improvement of the QHII data by future experiments it will be possible to reduce the error-bars in the reconstructed history and put further to the test the hyperbolic tangent modelling for reionization.\\
\begin{table*}
\begin{tabular}{ |p{2cm}||p{2.5cm}|p{2.5cm}|p{2.7cm}|p{2.7cm}|p{2.5cm}|p{2.5cm}|}
 \hline
 \multicolumn{7}{|c|}{B1} \\
 \hline
Parameters &Planck 2018+UV17& Planck 2018+QHII& Planck 2018+UV17+QHII &Planck 2018-SR2+UV17+QHII & Planck -TT +UV17+QHII & Planck Lens+UV17+QHII\\
 \hline
$\Omega_b h^2$ & $0.0224\pm 0.0001$& $0.0224\pm 0.0001$ & $0.0224\pm 0.0001$& $0.0224\pm 0.0001$&  $0.0221\pm 0.0002$&  -\\
$\Omega_c h^2$ &  $0.120\pm 0.001$& $0.120\pm 0.001$ & $0.120\pm 0.001$& $0.120\pm 0.001$& $0.120\pm 0.001$& $0.113\pm 0.008$\\
$100\theta_{MC}$ & $1.0409\pm 0.0003$& $1.0409\pm 0.0003$ & $1.0409\pm 0.0003$& $1.0409\pm 0.0003$& $1.0408\pm 0.0004$& -\\
$\tau$ &  $0.0544_{-0.0077}^{+0.0058}$&$0.05350_{-0.0039}^{+0.0019}$ & $0.0519_{-0.0008}^{+0.0010}$& $0.0521_{-0.0008}^{+0.0010}$ &  $0.0515_{-0.0009}^{+0.0011}$& $0.0532\pm 0.0015$\\
$z_{int}^1$ & $21.1_{-3.5}^{+6.8}$& $20.9_{-2.9}^{+8.9}$ & $21.2_{-3.5}^{+6.5}$& $21.2_{-3.6}^{+6.6}$ &  $21.1_{-3.3}^{+6.8}$ & $21.2_{-3.3}^{+6.8}$\\\
\makecell[l]{$\log_{10}[\rho_{\rm UV}]$\\\tiny{$[{\rm erg}/({\rm s}\, {\rm H_z} {\rm Mpc}^{3})]$}}& $21.0_{-2.5}^{+1.8}$& $21.9_{-3.2}^{+2.7}$ & $21.0_{-2.4}^{+1.8}$& $21.0_{-2.4}^{+1.8}$&$21.0_{-2.5}^{+1.7}$ &$21.0_{-2.6}^{+1.7}$\\
$\log_{10}[R^1]$ & $-$ & $-$ & $-$& $-$ & $-$& $-$\\
\makecell[l]{$\log_{10}[\dot{\xi}_{\rm ion}]$\\\tiny{$[{\rm erg\, H_z^{-1}}]$}} & $25.04_{-0.30}^{+0.23}$& $24.86_{-0.13}^{+0.19}$ & $24.95_{-0.05}^{+0.04}$& $24.95_{-0.04}^{+0.04}$ &  $24.94_{-0.03}^{+0.04}$ &$24.94_{-0.04}^{+0.05}$\\
${\rm{ln}}(10^{10} A_s)$ & $3.044_{-0.015}^{+0.012}$& $3.043_{-0.009}^{+0.007}$ & $3.040\pm{0.006}$& $3.041\pm{0.006}$ &$3.039\pm 0.006$ &$3.054\pm{0.019}$\\
$n_s$ &$0.9648\pm 0.0040$& $0.9650\pm{0.0038}$ & $0.9645\pm{0.0038}$& $0.9640_{0.0038}^{0.0039}$ &  $0.9628\pm 0.0004$ & $-$\\
$ \Delta_z^{\rm Reion}$ & $2.78_{-0.14}^{+0.12}$& $3.07_{-1.69}^{+0.37}$ & $ 2.80_{-0.15}^{+0.12}$& $2.81_{-0.16}^{+0.12}$ &$2.80_{-0.15}^{+0.12}$ &$-$\\
 \hline
\end{tabular}
\caption{~\label{Table3}Constraints on the cosmological and reionization parameters for the B1 case with the efficiency free to vary. The different columns represent different data combinations, the first three columns consider the Planck baseline CMB data whereas the last three columns represent the variation of the CMB data used. Error bars are the 68\% C.L. The $\Delta_z$ for the lensing only run is not derived due to the fixed cosmological parameters assumed.}
\end{table*}
\subsubsection{Varying the neutrino mass}
Within the discussion on the impact of the reionization reconstruction on the cosmological model we have considered an extended cosmological model whose constraints have shown a sensitivity to the reionization history, the case of a varying neutrino mass \cite{Archidiacono:2010wp,Archidiacono:2016lnv,HPFS18,Paoletti:2020ndu}. This model is particularly important being considered one of the baseline models for future large scale structure experiments whose combination with future CMB data will have the power to constrain the neutrino mass \cite{CORE:2016npo,Ballardini:2021frp}.

\begin{figure*}
\includegraphics[width=\textwidth]{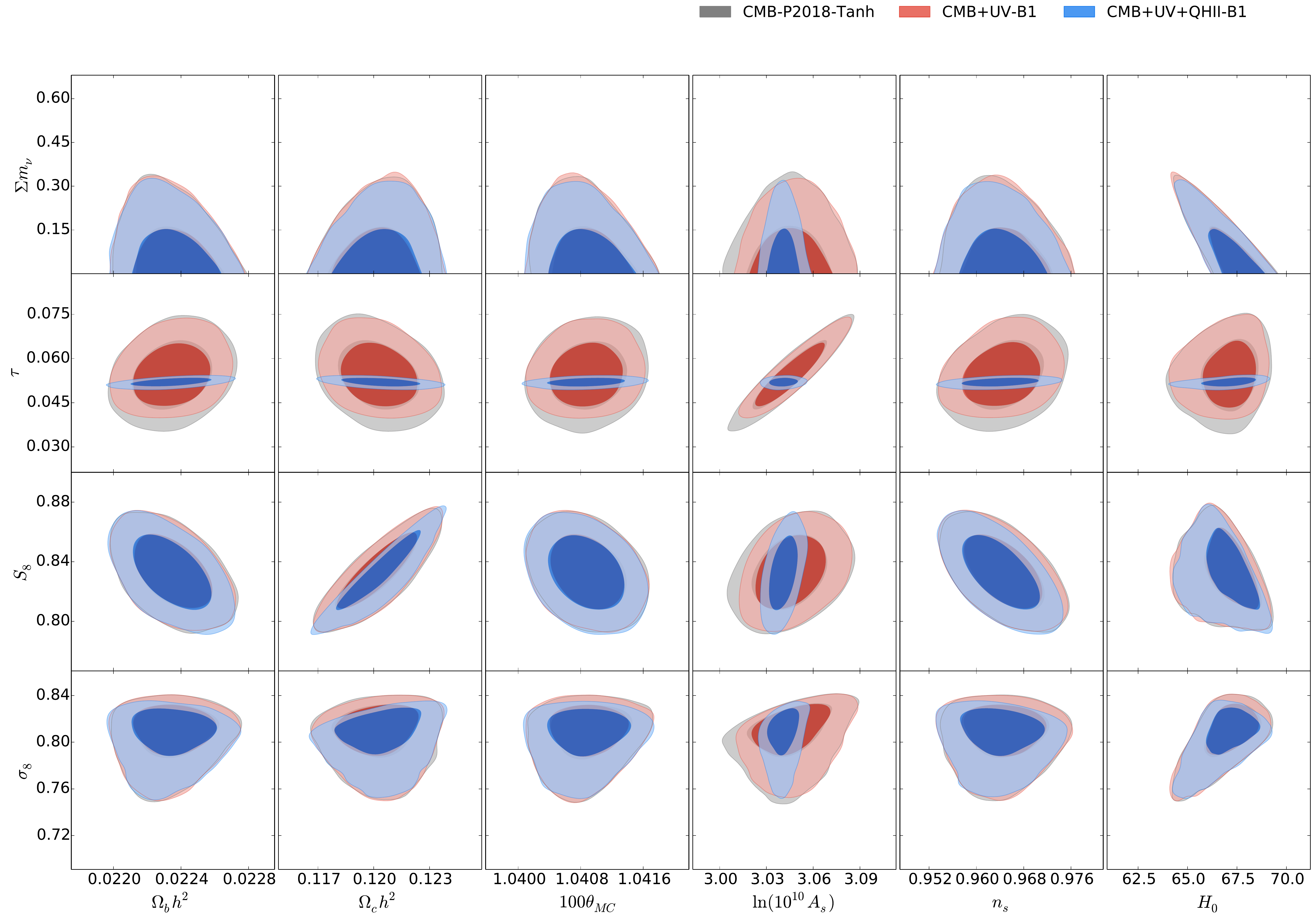}
\caption{~\label{MnuF}Two dimensional contours for the constraints on cosmological parameters for the varying neutrino mass case. The reconstruction is compared with the hyperbolic tangent case.}
\end{figure*}

If we additionally sample over the sum of neutrino masses, assuming a degenerate hierarchy, being current experiments not sensitive to the hierarchy, we obtain the results in~\autoref{MnuF} compared with the hyperbolic tangent case. For this specific model since we want to investigate specifically the impact of reionization we do not consider the Planck lensing likelihood which alone would be able to tighten the constraints on the neutrino mass desensitizing the analysis to the reionization impact. We note how the reconstruction does not affect the neutrino mass constraints and provides $\Sigma m_\nu<0.25\,,\tau=0.0520\pm 0.001$ with the full data combination and  $\Sigma m_\nu<0.26\,,\tau=0.0550_{-0.008}^{+0.006}$ for the case with only CMB+UV; for comparison the hyperbolic tangent provides $\Sigma m_\nu<0.26\,,\tau=0.0545_{-0.007}^{+0.008}$. The addition of astrophysical data reduces the uncertainties on the optical depth and the central value but only marginally affects the neutrino mass constraint.
We can therefore conclude that for current data the reconstruction does not affect the $\Lambda$CDM+neutrino mass model.
\subsubsection{Non-monotonic reionization}
\begin{figure}
\includegraphics[width=0.5\textwidth]{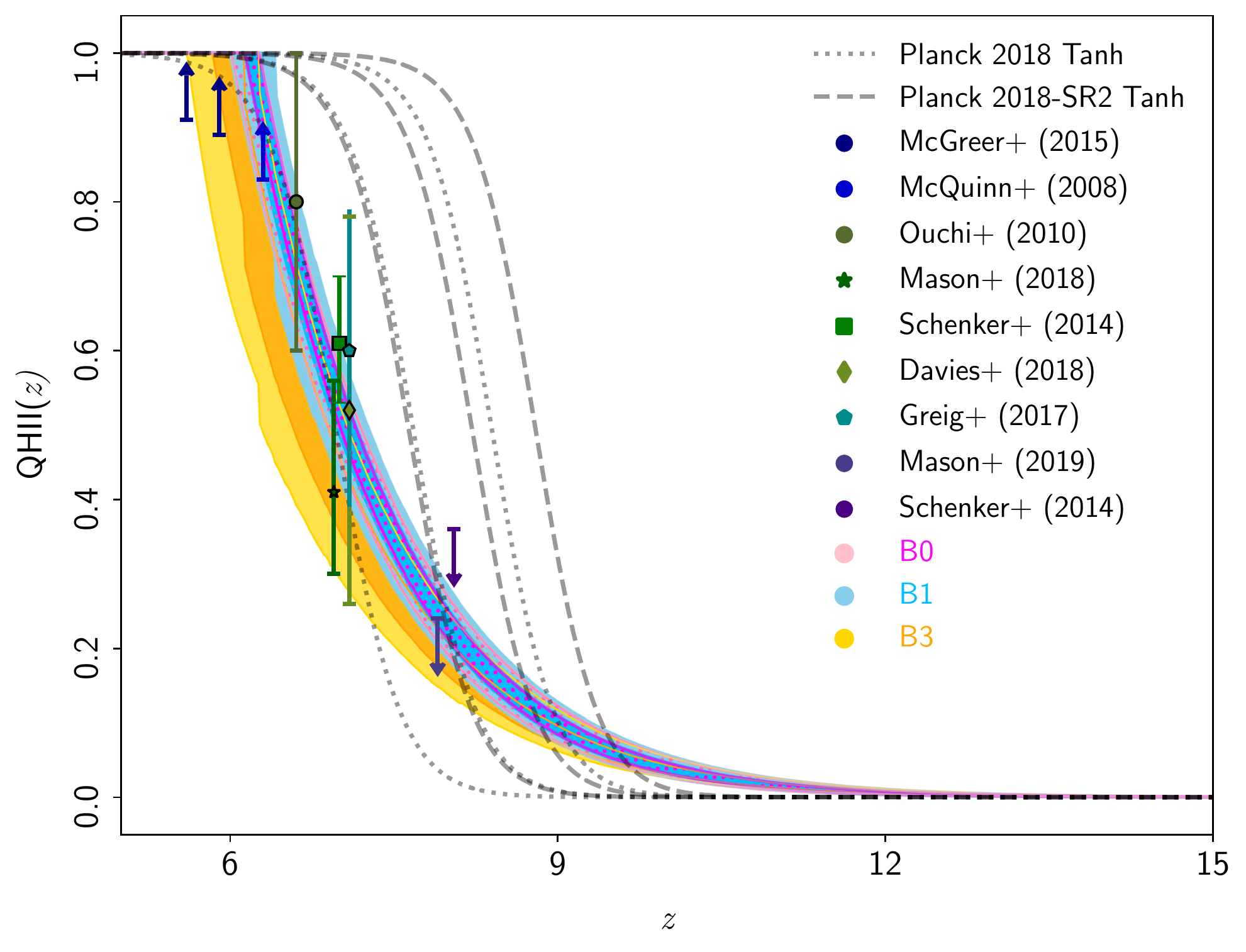}
\caption{~\label{Rec2}Reconstructed reionization histories using the full data combination of astrophysical data and CMB. We present the comparison for free $\dot{\xi}_{\rm ion}$ between the single node cases and the three node case. In gray lines the 68\% confidence level band for the hyperbolic tangent in Planck 2018 baseline results.}
\end{figure}
\begin{table}
\begin{tabular}{ |p{2.5cm}||p{4cm}| }
 \hline
 \multicolumn{2}{|c|}{B3} \\
 \hline
Parameters & Planck 2018+UV17+QHII \\ \hline
$\Omega_b h^2$ & $0.0224\pm{0.0001(0.0027)}$\\
$\Omega_c h^2$ & $0.1202\pm{0.001(0.002)}$\\
$100\theta_{MC}$ & $1.0409\pm{0.0003(0.0006)}$\\
$\tau$ & $0.0515_{-0.0019(-0.0033)}^{+0.0009(+0.0041)}$\\
$z_{int}^1$ & $-$\\
 $\log_{10}[\rho_{\rm UV}^2]$& $25.78_{-0.17(-0.20)}^{+0.08(+0.26)}$\\
$\log_{10}[R^1]$ & $<0.08$\\
$z_{int}^2$ & $-$\\
$ \log_{10}[\rho_{\rm UV}^2]$ & $24.67_{-0.35(-0.74)}^{+0.40(+0.73)}$\\
$\log_{10}[R^2]$ & $-$\\
$z_{int}^3$ & $-$\\
$ \log_{10}[\rho_{\rm UV}^3]$ & $20.4_{-4.0(-5.2)}^{+2.8(+4.8)}$\\
$\log_{10}[R^3]$ & $-$\\
\makecell[l]{$\log_{10}[\dot{\xi}_{\rm ion}]$\\\tiny{$[{\rm erg\, H_z^{-1}}]$}} & $24.94_{-0.08(-0.13)}^{+0.03(+0.18)}$\\
${\rm{ln}}(10^{10} A_s)$ & $3.039_{-0.007(-0.013)}^{+0.006(+0.013)}$\\
$n_s$ & $0.9644_{-0.0038(-0.0076)}^{+0.0039(0.0074)}$\\
$ \Delta_z^{\rm Reion}$ & $2.91_{-0.18}^{+0.07}$\\
\hline
\end{tabular}
\caption{~\label{TableB3}68\% C.L. constraints on the parameters for the B3 case.In parentheses we report the 95\% C.L. uncertainty, upper bounds are the 95\% C.L..}
\end{table}

We performed an exploratory analysis of the impact of the variation of the efficiency for non-monotonic histories of reionization considering three nodes. The results are reported in~\autoref{TableB3} we note that in this configuration the freedom is enough to not allow to constrain the node position and recombination time but thanks to the full data combination is possible to constrain the UV luminosity density and the efficiency $\log_{10}[\dot{\xi}_{\rm ion}]/[erg^{-1} H_z]=24.94_{-0.08,-0.13}^{+0.03,+0.18}$ which is again in agreement at 95\% C.L. with the assumption we made in~\cite{PRL}. We stress that non-monotonic histories of reionization are anyway already strongly disfavoured by current data and the present analysis is only meant to confirm the robustness of our treatment against reconstructions with higher degrees of freedom. 

In ~\autoref{Rec2} we present the reconstructed reionization histories using the full combination of CMB and astrophysical data comparing the three cases of B0, B1 and B3 with varying efficiency. The histories are perfectly compatible with the case B0 superimposed to the B1. The case with three nodes enlarges the error bars as expected and allows for a later reionization. The B3 case the lowest redshift node is strongly sensitive to the QHII data whereas the intermediate node is more sensitive to the UV data, this increased freedom weights more on the lower redshift data by the QHII slightly moving the reionization towards later times.
In this three bins case we have also derived the value of the optical depth limited only to high redshift which in this case means $z>15$. for this test we keep fixed the efficiency since we already shown that the high redshift tail is not impacted by this parameter. The results provide $\tau_{z>15}<0.001$ which is below the 68\% C.L. error bar. We can conclude that the contribution to the reionization history for redshift higher than 15 is negligible.

\subsection{Changing the data}

\subsubsection{Low-tail of luminosity function}

The luminosity function used to derive the UV luminosity density has a non-trivial shape collecting different contributions. 
In particular, in the derivation of the UV luminosity density we have assumed a conservative truncation magnitude of -17. This choice cuts the low-luminosity tail of the function where indeed we do expect significant contribution by fainter, but more abundant, sources. The issue of the low-luminosity tail concerns the behaviour of the current data which is still not clearly determined. As shown in~\cite{Ishigaki2018} the faint end seems to show a possible acceleration of the curve , in the data compilation we used, which is still not consolidated by data but may have an impact on the UV luminosity density. For this reason in our previous work we assumed a conservative cut.

We now investigate the impact of choosing a less conservative assumption in the derivation of the UV luminosity density by using a truncation magnitude of -15 including therefore the low-luminosity tail in our analysis. Also for this analysis we leave free to vary the ionizing efficiency term of the source.

\begin{table}
\begin{tabular}{ |p{2.3cm}||p{3cm}||p{3cm}|}
 \hline
 \multicolumn{3}{|c|}{B0} \\
 \hline
Parameters & P2018+UV15 & P2018+UV15+QHII \\ \hline
$\Omega_b h^2$ & $0.0224\pm 0.0001$& $0.0224\pm 0.0001$\\
$\Omega_c h^2$ & $0.120\pm 0.001$& $0.120\pm 0.001$\\
$100\theta_{MC}$ & $1.0409\pm 0.0003$& $1.0409\pm 0.0003$\\
$\tau$ & $0.0550_{-0.0081}^{+0.0055}$& $0.0542_{-0.0016}^{+0.0014}$\\
$z_{int}^1$ & $21.1_{-2.9}^{+8.0}$& $21.1_{-3.1}^{+7.8}$\\
\makecell[l]{$\log_{10}[\rho_{\rm UV}]$\\\tiny{$[{\rm erg}/({\rm s}\, {\rm H_z} {\rm Mpc}^{3})]$}}& $23.2_{-1.5}^{+0.8}$& $23.2_{-1.4}^{+0.9}$\\
\makecell[l]{$\log_{10}[\dot{\xi}_{\rm ion}]$\\\tiny{$[{\rm erg\, H_z^{-1}}]$}} & $24.72_{-0.26}^{+0.19}$& $24.70_{-0.05}^{+0.06}$\\
${\rm{ln}}(10^{10}A_s)$ & $3.046_{-0.015}^{+0.011}$& $3.044\pm 0.006$\\
$n_s$ & $0.9650\pm 0.0040$& $0.9649\pm 0.0038$\\
$ \Delta_z^{\rm Reion}$ & $4.21_{-0.70}^{+0.33}$ & $4.20_{-0.64}^{+0.36}$ \\
\hline
\end{tabular}
\caption{~\label{TableB015}Constraints on the parameters for the B0 case using the UV luminosity density cut at -15 in the luminosity function. Error bars are the 68\% C.L.}
\end{table}

\begin{table}
\begin{tabular}{ |p{2.3cm}||p{2.5cm}||p{3.5cm}|}
 \hline
 \multicolumn{3}{|c|}{B1} \\
 \hline
Parameters & P2018+UV15 & P2018+UV15+QHII {\it{(SROLL2)}} \\ \hline
$\Omega_b h^2$ & $0.0224\pm 0.0001$& \makecell{$0.0224\pm 0.0001$ \\ $(0.0224\pm 0.0001)$}\\
$\Omega_c h^2$ & $0.120\pm 0.001$&\makecell{ $0.120\pm 0.001$\\ ($0.120\pm 0.001)$}\\
$100\theta_{MC}$ & $1.0409\pm 0.0003$& \makecell{$1.0409\pm 0.0003$\\$(1.0409\pm 0.0003)$}\\
$\tau$ & $0.0549_{-0.0077}^{+0.0057}$& \makecell{$0.0541_{-0.0016}^{+0.0013}$\\$(0.0544_{-0.0016}^{+0.0014})$}\\
$z_{int}^1$ & $20.7_{-3.2}^{+8.5}$& \makecell{$21.3_{-3.0}^{+7.7}$\\($21.8_{-2.8}^{+7.4}$)}\\
\makecell[l]{$\log_{10}[\rho_{\rm UV}]$\\\tiny{$[{\rm erg}/({\rm s}\, {\rm H_z} {\rm Mpc}^{3})]$}}& $23.3_{-1.6}^{+0.9}$& \makecell{$23.1_{-1.4}^{+0.9}$\\$(23.1_{-1.3}^{+0.9})$}\\
$\log_{10}[R^1]$ & $-$& \makecell{$-$\\$(-)$}\\
\makecell[l]{$\log_{10}[\dot{\xi}_{\rm ion}]$\\\tiny{$[{\rm erg\, H_z^{-1}}]$}} & $24.78_{-0.34}^{+0.16}$& \makecell{$24.71\pm 0.06 $\\$(24.70_{-0.05}^{+0.06})$}\\
${\rm{ln}}(10^{10}A_s)$ & $3.045_{-0.015}^{+0.012}$& \makecell{$3.044\pm 0.006$ \\$(3.045\pm0.006)$}\\
$n_s$ & $0.9649\pm 0.0040$& \makecell{$0.9647\pm 0.0040$\\ $(0.9645\pm 0.0040)$}\\
$ \Delta_z^{\rm Reion}$ & $4.19_{-0.71}^{+0.31}$ & \makecell{$4.22_{-0.62}^{+0.33}$\\ $(4.30_{-0.65}^{+0.38})$}\\
\hline
\end{tabular}

\caption{~\label{TableB115}Constraints on the parameters for the B1 case using the UV luminosity density cut at -15 in the luminosity function. As before the error bars are 68\% C.L. The numbers in parentheses are the results with SROLL2 likelihood.}
\end{table}

\begin{figure}
\includegraphics[width=0.5\textwidth]{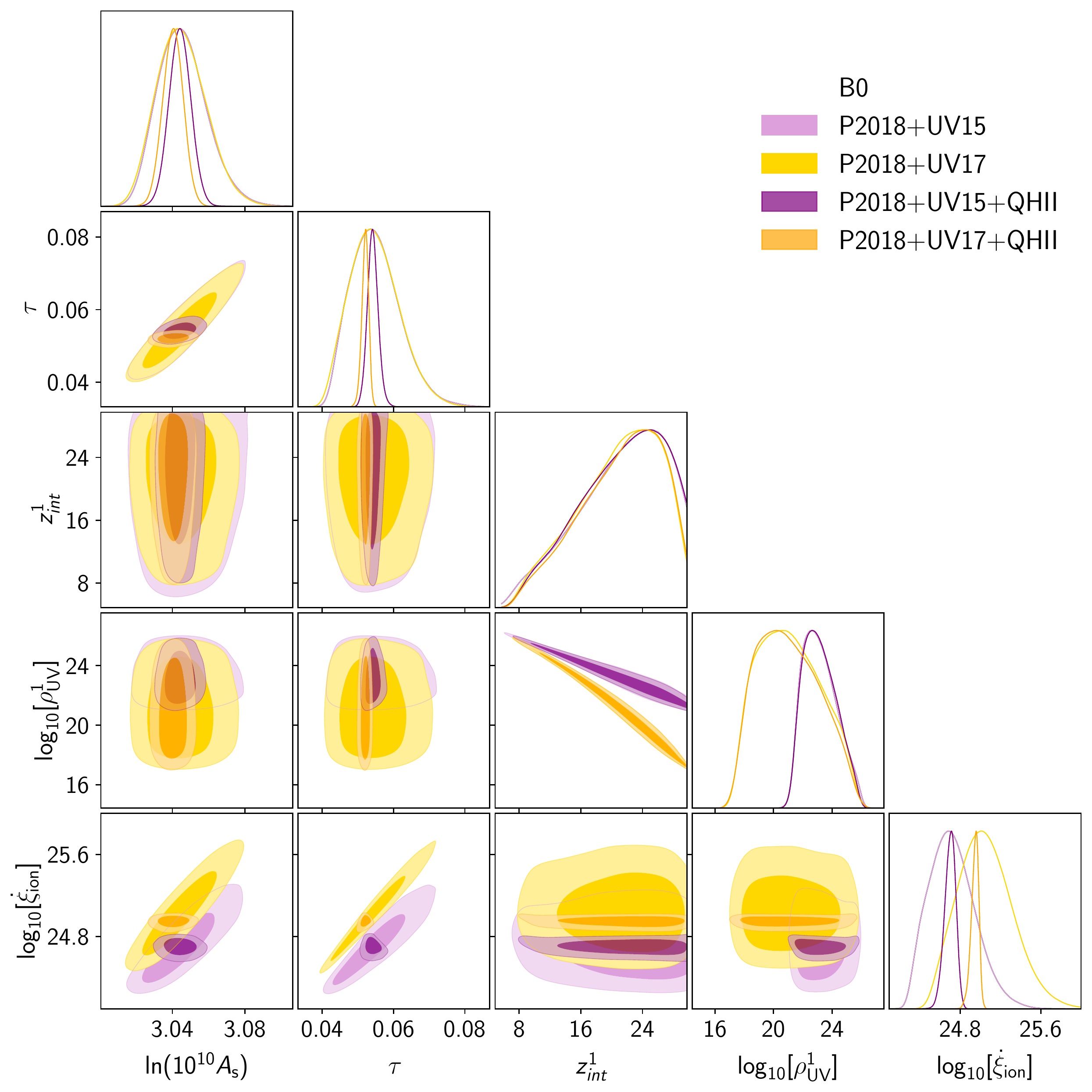}
\caption{~\label{F151}Two dimensional contours for the constraints on the reionization parameters using the luminosity function magnitude cut at -15 for the B0 case compared with the same data combination with -17 cut.}
\end{figure}

\begin{figure}
\includegraphics[width=0.5\textwidth]{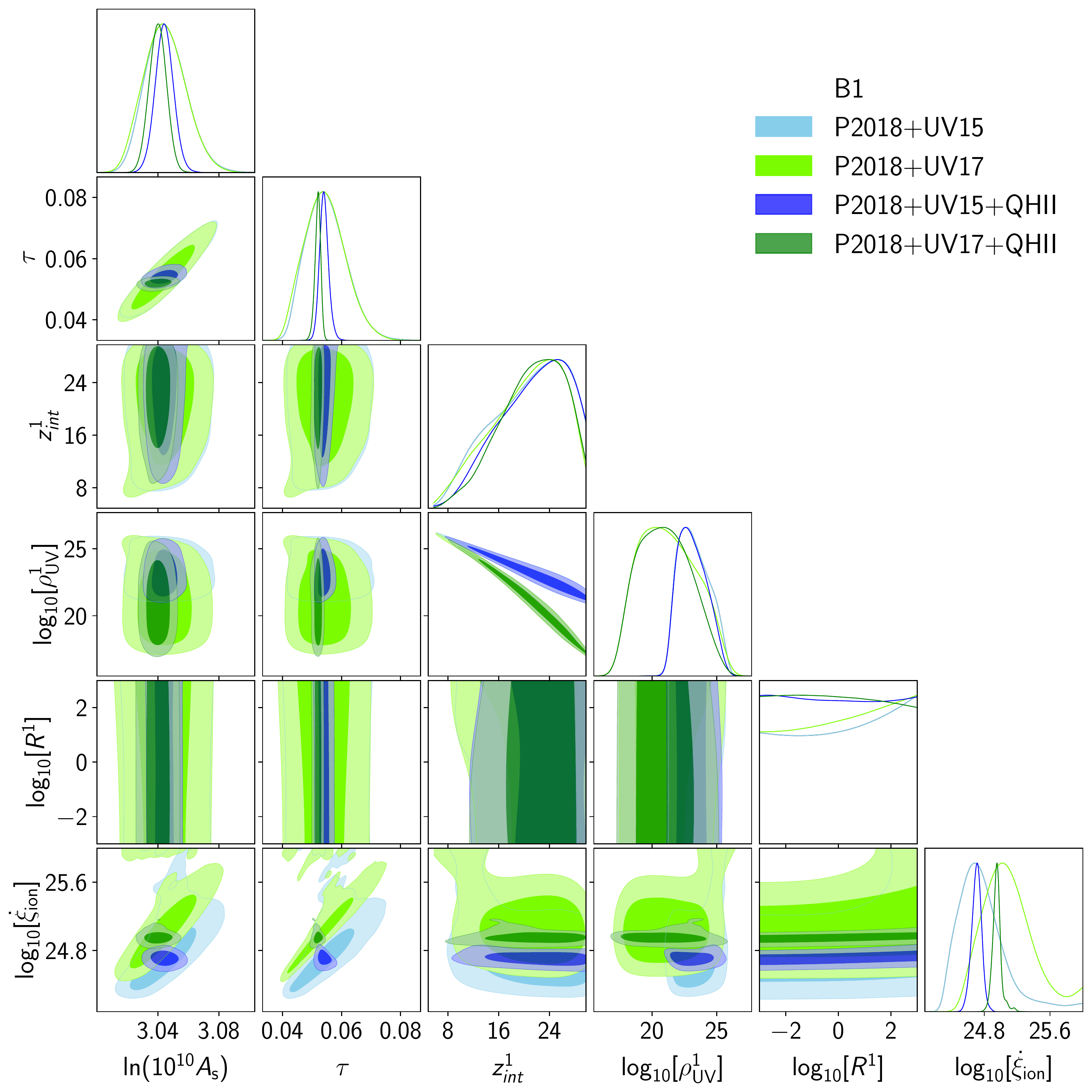}
\caption{~\label{F152}Two dimensional contours for the constraints on the reionization parameters using the luminosity function magnitude cut at -15 for the B1 case compared with the same data combination using the -17 cut.}
\end{figure}

The results are shown in~\autoref{F151} for the monotonic single burst case B0 whose constraints are also presented in~\autoref{TableB015} and in~\autoref{F152} for the monotonic B1 case whose constraints are presented in~\autoref{TableB115}.

\begin{figure}
\includegraphics[width=0.5\textwidth]{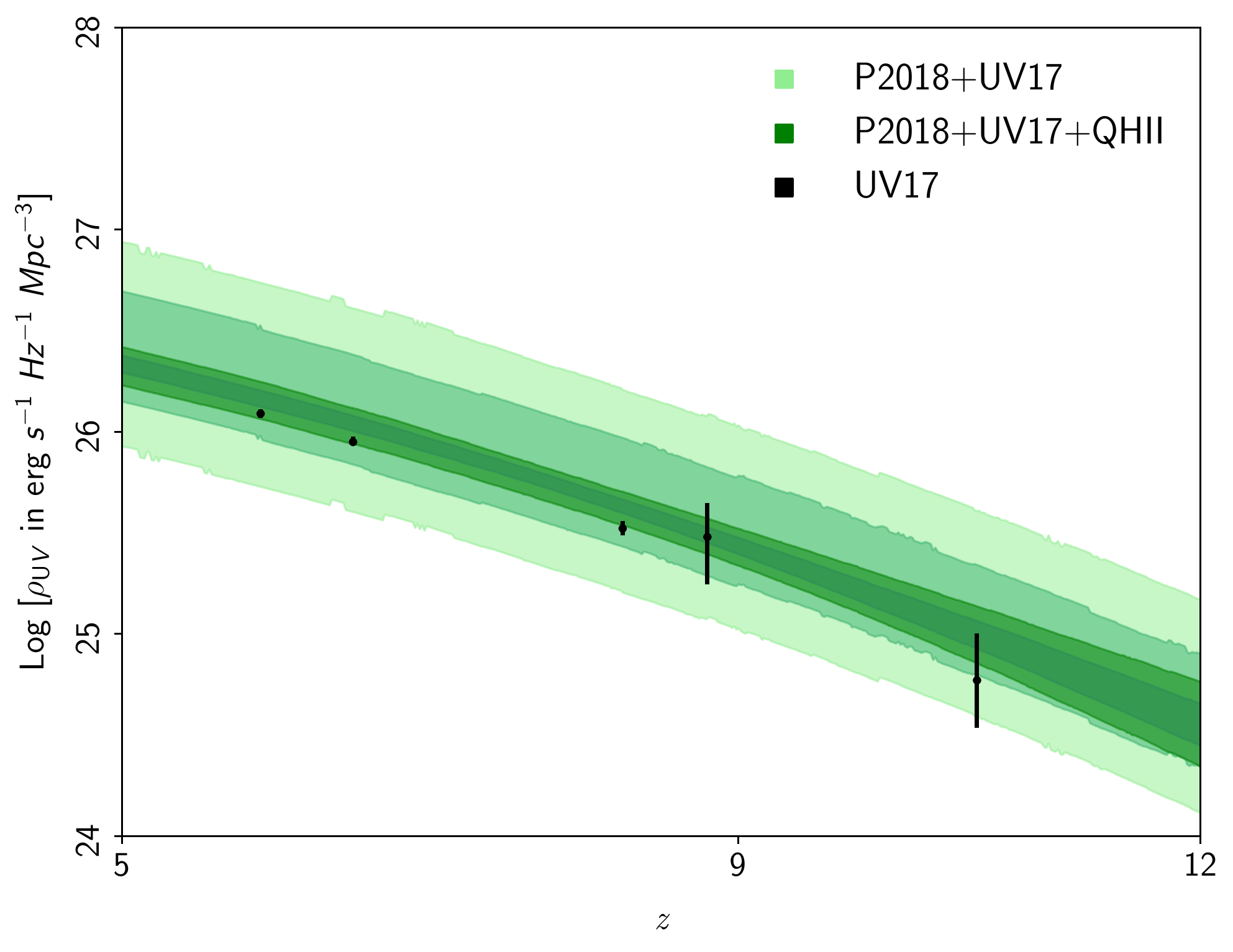}\\
\includegraphics[width=0.5\textwidth]{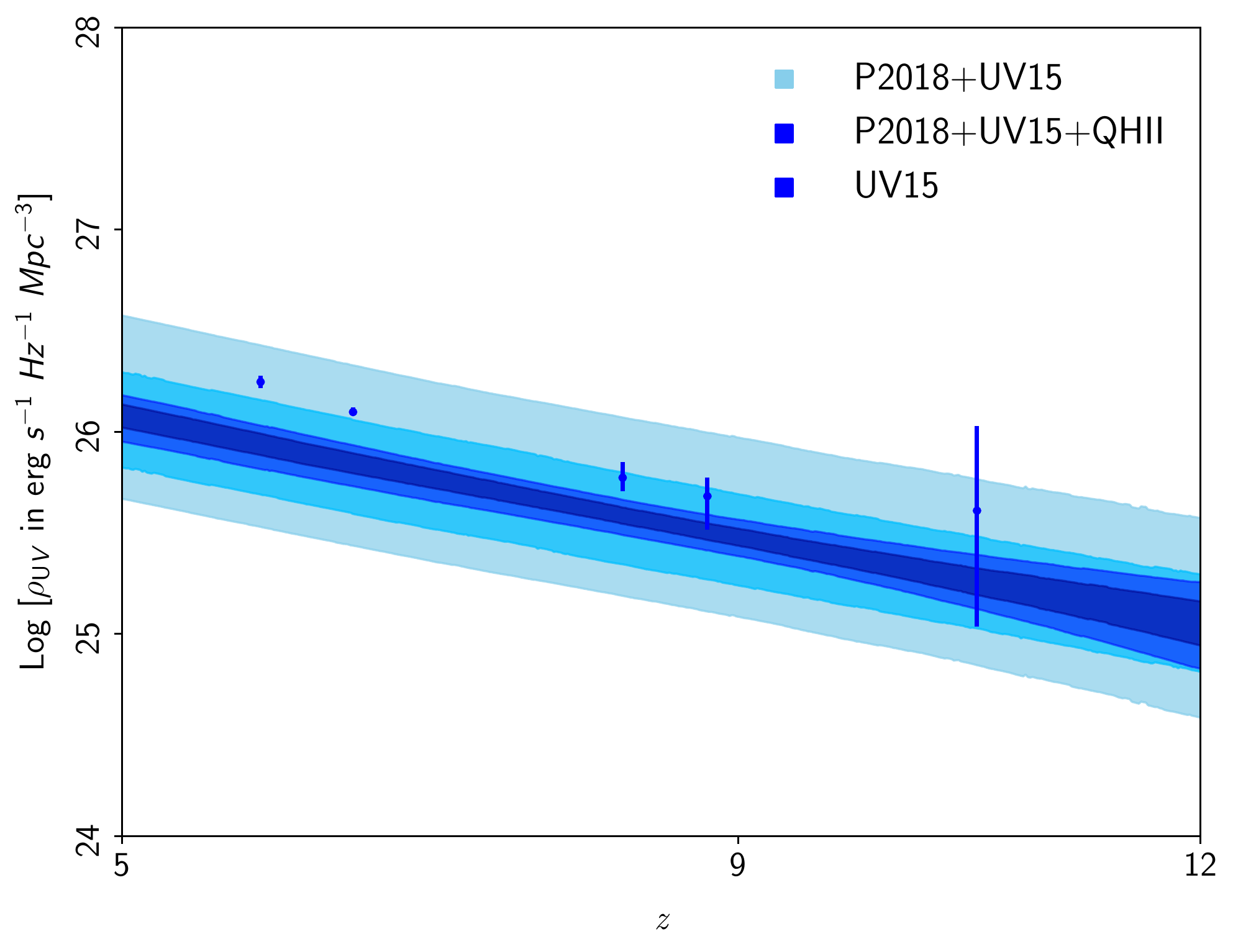}
\caption{~\label{F15UV}Reconstructed UV luminosity densities compared with the data for the full combination of Planck and astrophysical data. In green in the upper panel the case using the -17 cut in blue in the lower panel the case with the -15.}
\end{figure}

\begin{figure}
\includegraphics[width=0.5\textwidth]{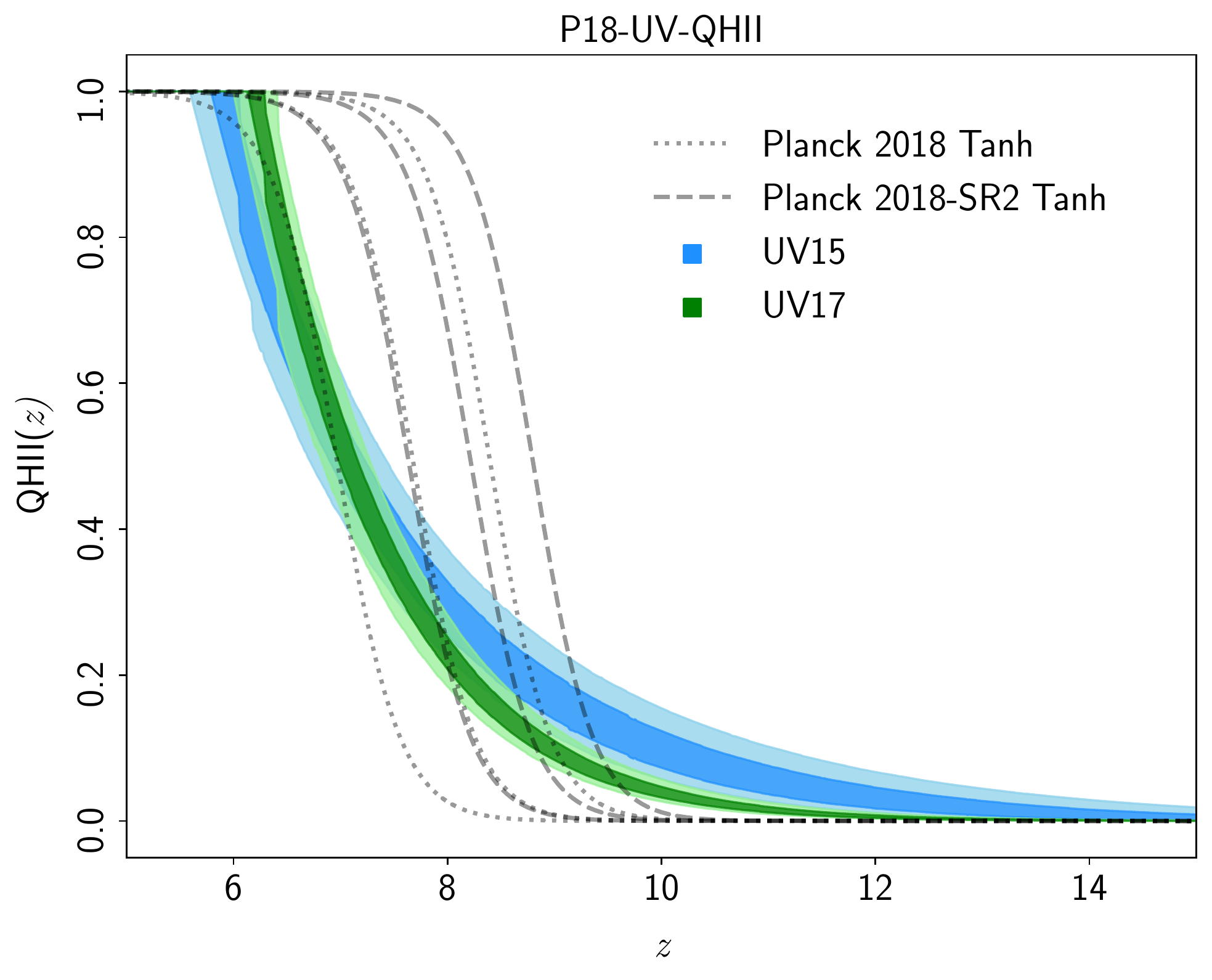}
\caption{~\label{F151Re}The reconstructed reionization histories for the B1 case with full data combination using either the -17 or the -15 magnitude cut in the UV luminosity function in addition to CMB+QHII.}
\end{figure}

From the point of view of cosmological parameters the use of -15 truncation magnitude does not impact the results for both configurations B0 and B1. The result is robust with respect to the addition of the QHII data which tightens the constraints. The only substantial difference is in the parameters of the reionization. The fainter magnitude cut prefers an higher optical depth of $\tau=0.0541^{+0.0013}_{-0.0016}$ as expected from the addition of the faint sources contribution which brings the central value towards the Planck 2018 hyperbolic tangent result.
The efficiency is pushed towards lower values by the more optimistic cut and some shift is observed in the reconstructed UV luminosity density and as a consequence in the efficiency component of the source term. These differences are expected due to the intrinsically different treatment of the UV luminosity density. 
In ~\autoref{F15UV} we show the reconstructed UV luminosity densities for the two cases compared with the data points used.The case with the combination of CMB and UV is fully compatible with the data. The addition of the QHII tightens the curve and tilts it. We note the two different tilts between the cut at -17 and at -15 which is due to the change in the data at the higher redshifts.
In~\autoref{F151Re} we show the comparison between the reconstructed reionization histories.
Although  the two 95\% C.L. partially overlaps in the central redshift region we note that the two reionization histories are substantially different. The addition of the faint end of the luminosity function increases the contribution from higher redshift sources moving towards a shallower reionization history. This earlier onset of reionization is in agreement with simulations including higher redshift sources of reionization such as the models with a pop III stars contribution \cite{Miranda:2016trf}. The minor steepness of the curve makes the reconstructed reionization history even in more disagreement with the hyperbolic tangent used in the standard cosmological analyses although the reconstructed optical depth is more in agreement with the Planck 2018 baseline results. This is mainly due to effect at the power spectrum level where in the E mode polarization this contribution is indeed much closer to the Planck best fit. Future CMB sensitivities in the E-mode may be able to distinguish among the effects of the two cuts.

\subsubsection{Selection of ionized fraction data}

After the investigation of different alternatives for the UV luminosity density we now consider the case of different selections for the QHII data from high redshift objects.
\begin{figure}
\includegraphics[width=0.5\textwidth]{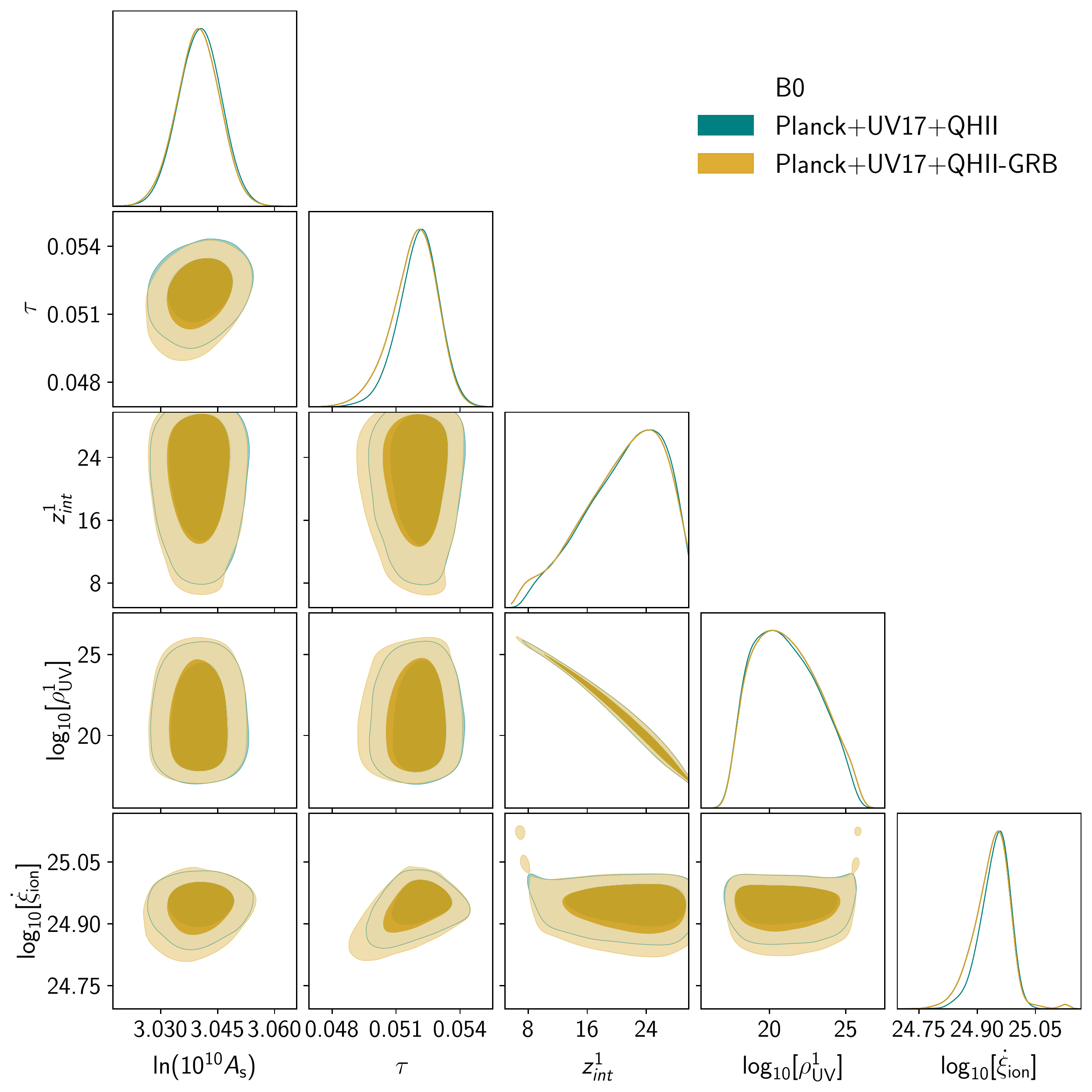}
\caption{~\label{GRB}Constraints on the reionization parameters from the combination of CMB, UV and QHII data excluding the GRB  050904 point from the QHII. We show here the B0 case.}
\end{figure}
%\begin{figure}
%\includegraphics[width=0.5\textwidth]{Tri_ReioB02018_BOGRB.pdf}\includegraphics[width=0.5\textwidth]{Tri_ReioB12018_BOGRB.pdf}
%\caption{~\label{GRB}Constraints on the reionization parameters from the combination of CMB, UV and QHII data excluding the GRB  050904 point from the QHII. On the left panel the B0 case on the right the B1 case.}
%\end{figure}
We start by cornering the effect of the only point we have which is based on a Gamma Ray Burst measurement and in particular on the damping wings of the Gamma Ray Burst 050904~\citep{Totani:2005ng,McQuinn:2007dy}.
This point having a different source for the data has a different process of data analysis and in the light of future Gamma Ray Burst data is interesting to see if it provides a different pull with respect to the more commonly used Quasar points. 
In~\autoref{GRB}, we show the comparison for  the monotonic single burst cases of the analysis excluding the Gamma Ray Burst point and the standard case which considers it and the monotonic case B1 provides analogous results. 
We do not note any appreciable difference between the two cases concluding that the GRB point addition does not impact the results but for a minimal reduction of the error bars it will be interesting to see if this reduction is increased by increasing the number of GRBs as expected from future data.

\begin{figure}
\includegraphics[width=0.5\textwidth]{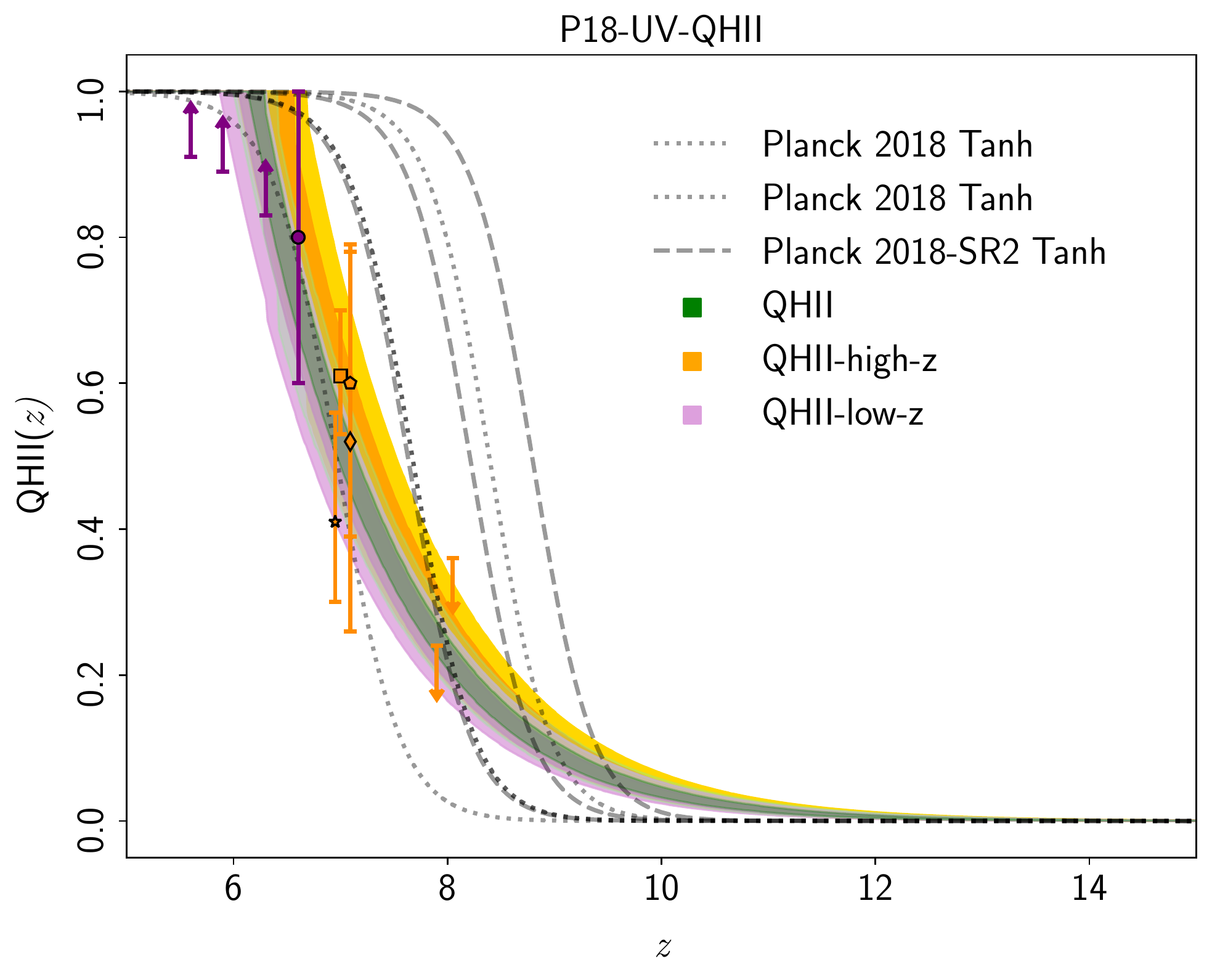}
\caption{~\label{F151HL} The reconstructed reionization histories using low or high redshift QHII data in addition to the CMB+UV.}
\end{figure}
Concerning the Quasar data we tested a series of different alternatives for the QHII points that were provided by slightly different data analysis pipeline. We divide the different points in three combinations named A1: as in~\citep{McGreer:2014qwa} we use a broader range in redshift as $[6.1, 0.62]$ instead of $[5.9, 0.94]$; A2: we use the alternative~\citep{Greig:2016vpu} point at $z=7.54$ using $0.79$ instead of $0.4$; A3:we use the~\citep{Davies:2018pdw} instead of the~\citep{Greig:2016vpu} point at $z=7.09$. 
The results for both the monotonic single burst and the monotonic cases do not show any appreciable difference. The alternatives to the data points we used do not modify our results and again confirm the  the robustness of our approach.

\begin{figure}
\includegraphics[width=0.5\textwidth]{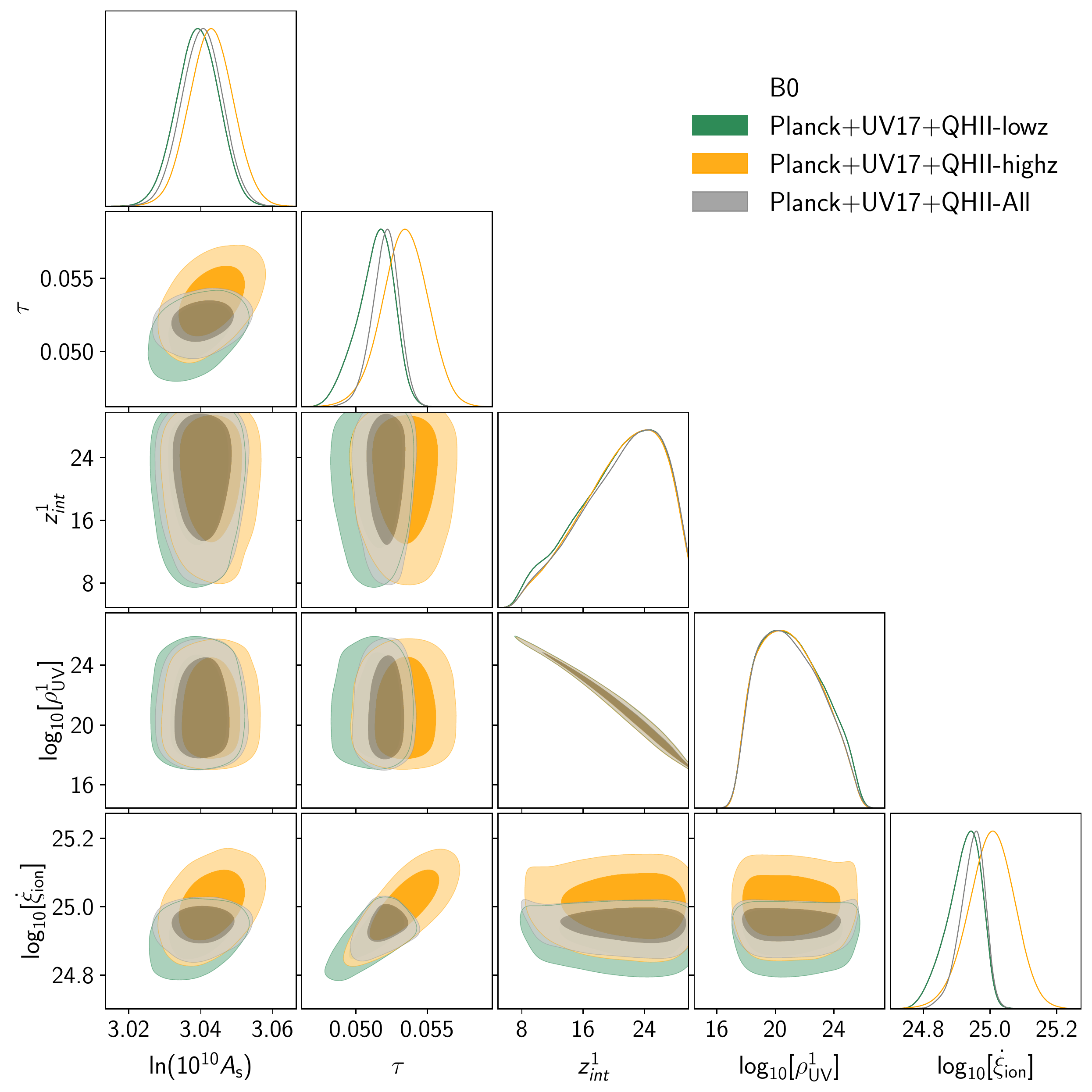}
\caption{~\label{lowhigh1}Constraints on the reionization parameters from the combination of CMB, UV and QHII data excluding the low redshift data and the high redshift data in turns. This is the B0 case.}
\end{figure}

\begin{figure}
\includegraphics[width=0.5\textwidth]{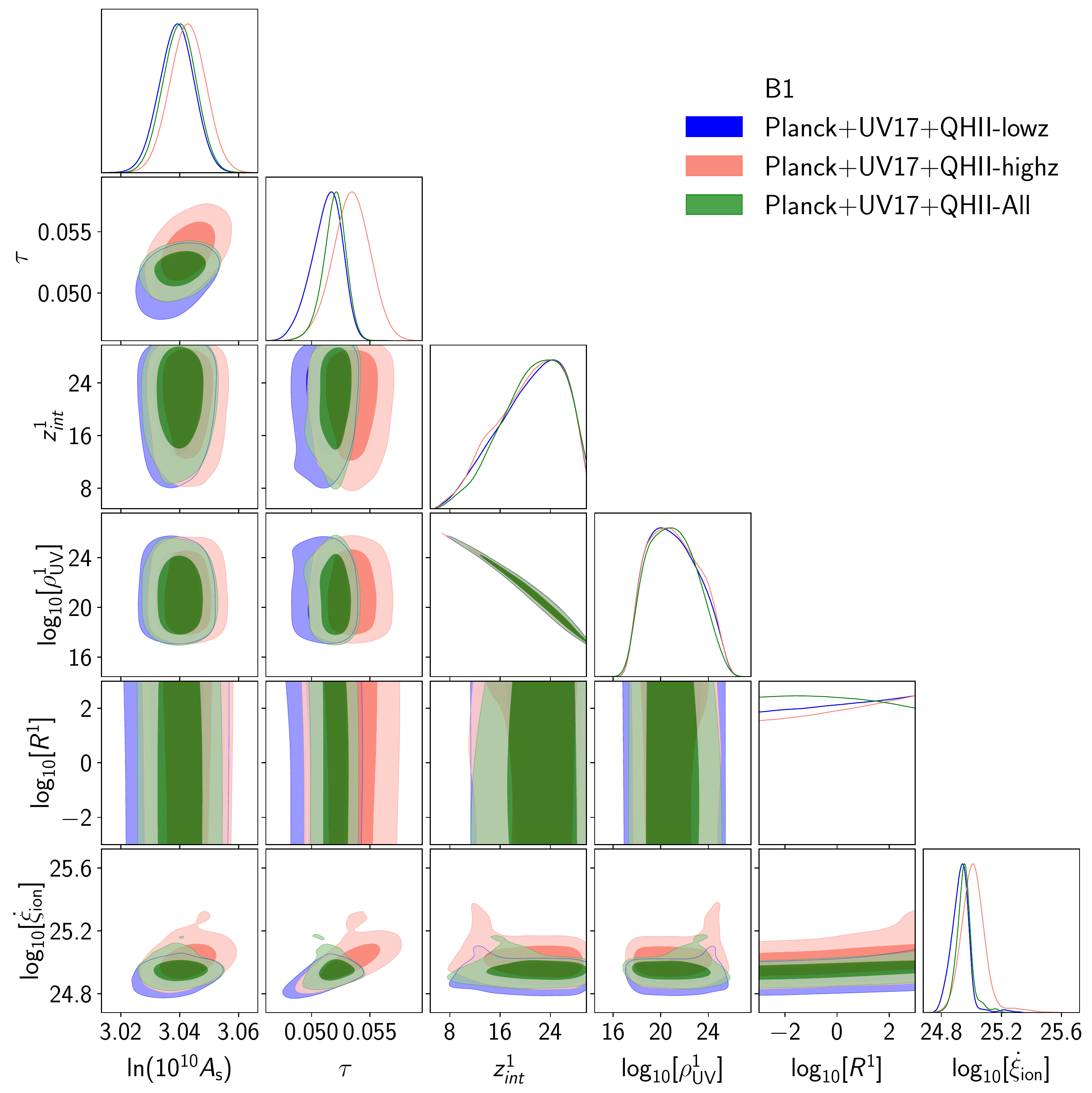}
\caption{~\label{lowhigh2}Constraints on the reionization parameters from the combination of CMB, UV and QHII data excluding the low redshift data and the high redshift data in turns. This is the B1 case.}
\end{figure}

We finally investigate the different pulls of QHII data at low redshifts ($z < 7$, denoted as low-$z$)
and high redshifts ($z > 7$, denoted as high-$z$). We display the reconstructed reionization
histories for the B1 case in the right panel of~\autoref{F151HL} and the two dimensional posteriors for the reionization
parameters for B0 and B1 in ~\autoref{lowhigh1} and ~\autoref{lowhigh2}. Cosmological parameters are in ~\autoref{TableB1HL}. Although perfectly consistent with the full dataset, also thanks to enlarged error bars, the two separate data sections show slighlty different pulls on the parameters. Higher redshift quasars prefer an higher value of the reconstructed optical depth and shift towards higher values of the efficiency. Lower redshift dataset instead shows the opposite behaviour.Again we note how the introduction of higher redshift data tends to move the reionization history towards higher optical depth and  earlier reionizations. 

\begin{table}
\begin{tabular}{ |p{2.3cm}||p{3cm}||p{3cm}|}
 \hline
 \multicolumn{3}{|c|}{B1} \\
 \hline
Parameters & P2018+UV17+QHII-low & P2018+UV17+QHII-high \\ \hline
$\Omega_b h^2$ & $0.0224\pm {0.0001}$& $0.0224\pm {0.0001}$\\
$\Omega_c h^2$ & $0.120\pm {0.001}$& $0.120\pm {0.001}$\\
$100\theta_{MC}$ & $1.0409\pm {0.0003}$& $1.0409\pm {0.0003}$\\
$\tau$ & $0.0513_{-0.0010}^{+0.0015}$& $0.0534_{-0.0016}^{+0.0016}$\\
$z_{int}^1$ & $20.9_{-3.4}^{+7.1}$& $20.7_{-3.5}^{+7.2}$\\
\makecell[l]{$\log_{10}[\rho_{\rm UV}]$\\\tiny{$[{\rm erg}/({\rm s}\, {\rm H_z} {\rm Mpc}^{3})]$}}& $21.1_{-2.7}^{+1.7}$& $21.2_{-2.7}^{+1.8}$\\
$\log_{10}[R^1]$ & $-$&$-$\\
\makecell[l]{$\log_{10}[\dot{\xi}_{\rm ion}]$\\\tiny{$[{\rm erg\, H_z^{-1}}]$}}  & $24.93_{-0.05}^{+0.06}$& $25.01_{-0.08}^{+0.06}$\\
${\rm{ln}}(10^{10}A_s)$ & $3.039\pm{0.006}$& $3.043\pm{0.006}$\\
$n_s$ & $0.9642\pm{0.0040}$& $0.9647\pm{0.0040}$\\
$ \Delta_z^{\rm Reion}$ & $2.79_{-0.14}^{+0.11}$ & $2.78_{-0.15}^{+0.12}$\\
\hline
\end{tabular}
\caption{~\label{TableB1HL}Constraints on the parameters for the B1 case using the low and high QHII selections. Error bars are the 68\% C.L.}
\end{table}
\section{Conclusions}

In this paper we have extended our previous analysis \cite{PRL} by considering different data and their combinations and by removing the assumption of a free ionization efficiency sampling the product $f_{\rm esc}\xi_{\rm ion}$ as a free parameter. \\
\begin{figure*}
\includegraphics[width=0.32\textwidth]{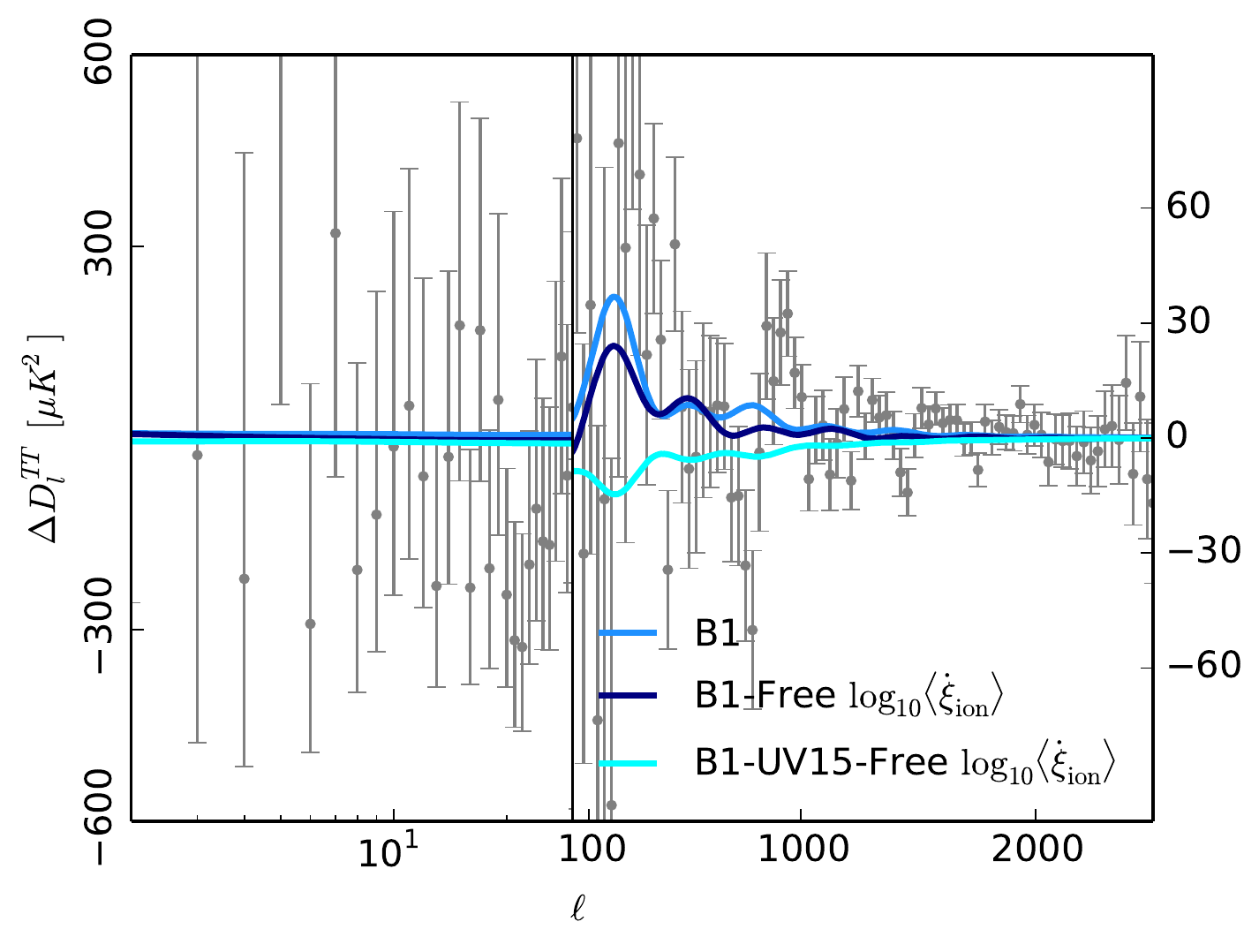}
\includegraphics[width=0.3\textwidth]{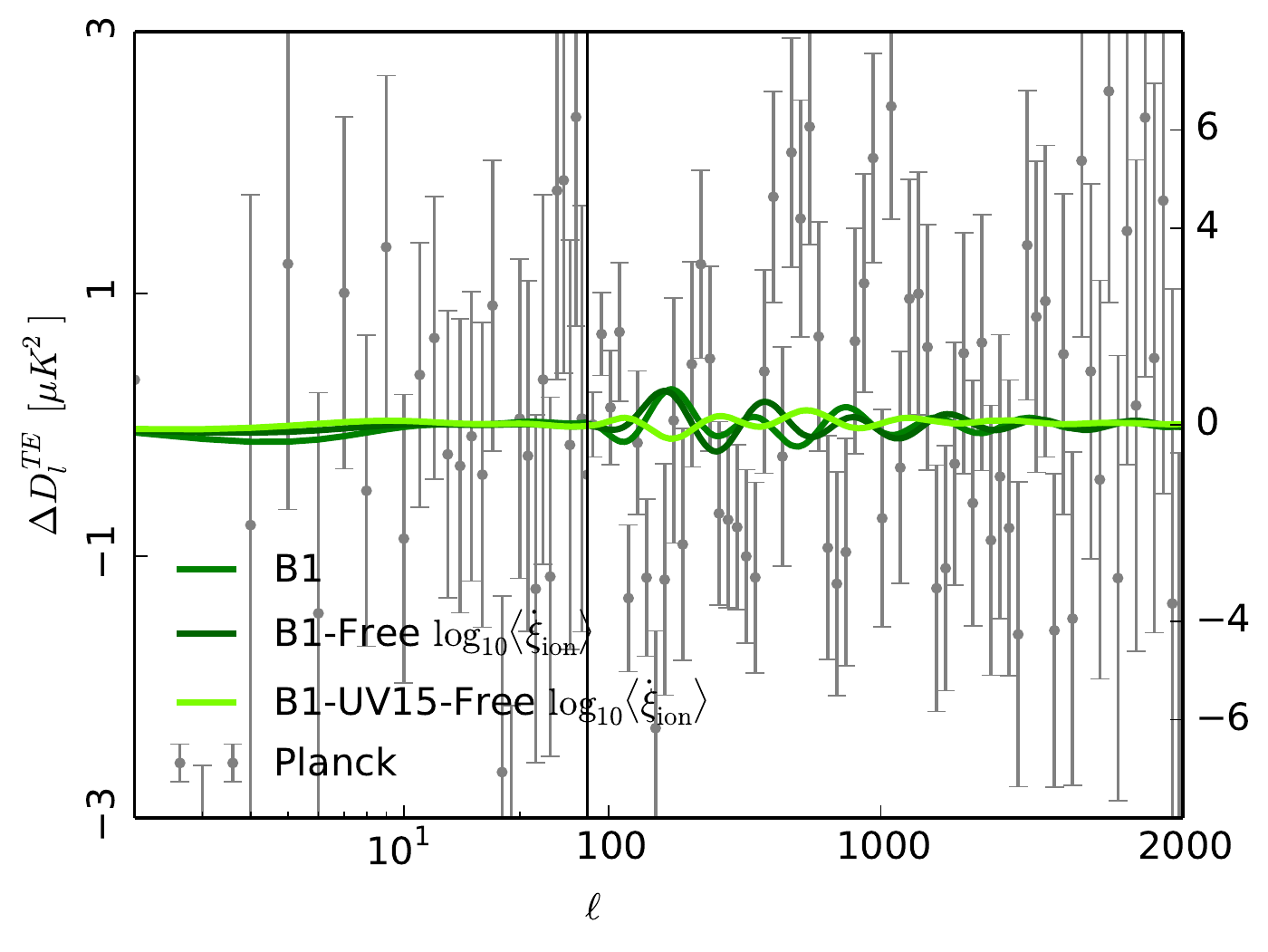}
\includegraphics[width=0.3\textwidth]{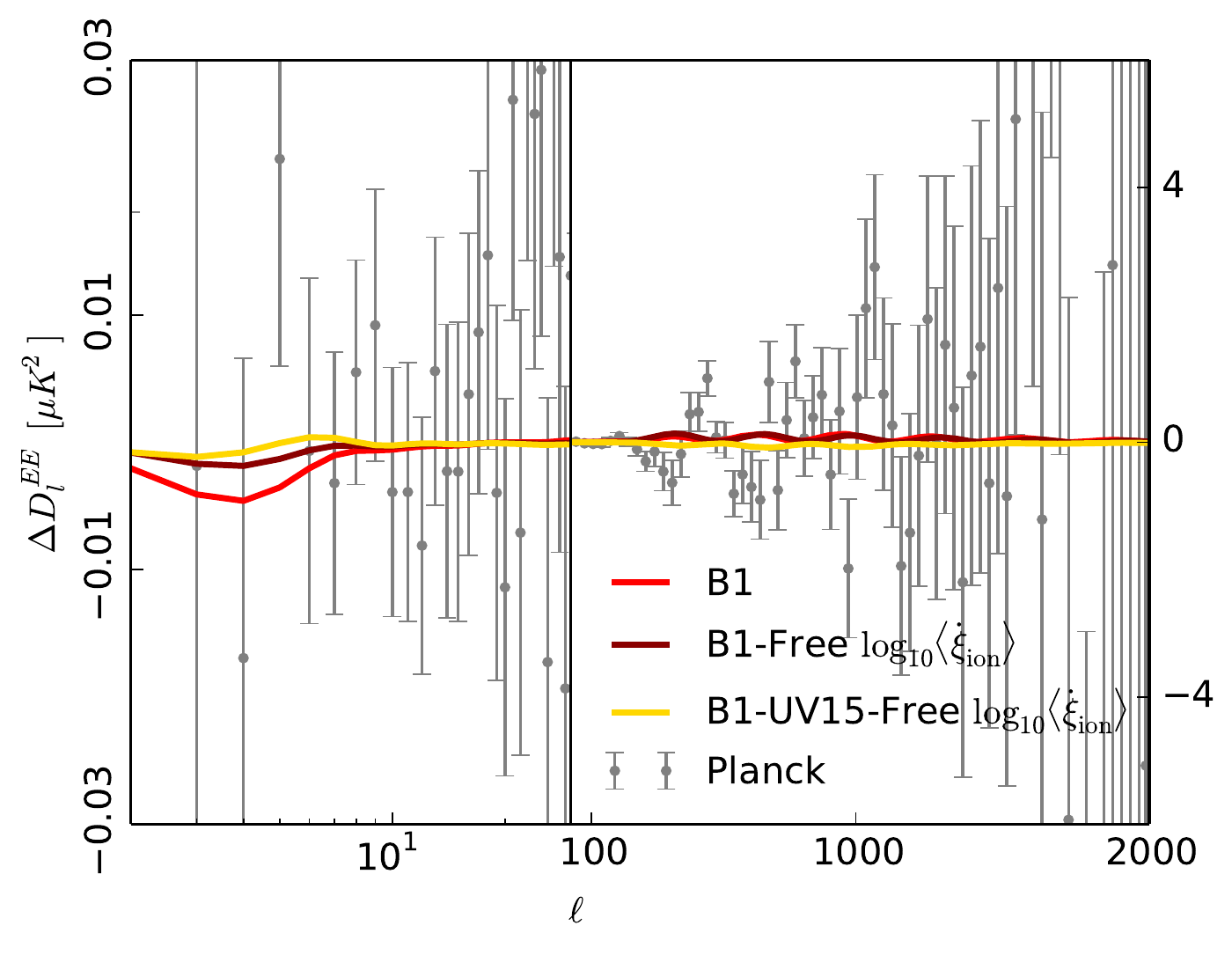}
\caption{Residuals with respect to Planck best fit, with hyperbolic tangent, of the B1 case best fit compared with the residuals from B1 with free efficiency and with the UV 15 cut. In gray are the   the data points from Planck. Left panel is TT, middle panel is TE and right panel is EE.}
\label{SpectPost}
\end{figure*}
When allowing to vary the ionization efficiency the results for both single node, either considering a single burst or the possibility of step like histories, and three node reconstruction show an excellent agreement with our previous results. The optical depth constraint is slightly affected by the free ionizing efficiency, the baseline B1 case providing $\tau=0.0519^{+0.0010}_{-0.0008}$, which is slightly larger than our previous result for fixed efficiency. CMB alone is not able to provide constraints on the efficiency, the addition of UV data improves but with very large uncertainties. The tightest constraint on the efficiency comes from the addition of the QHII data, which drastically reduces the uncertainties and flips the degeneracy direction of the efficiency with the optical depth and the scalar fluctuations amplitude leading to stronger constraints. The recovered ionization efficiency constraints are consistent within 95\% C.L. with the value assumed in \cite{PRL} for B1. The reconstructed ionization histories are in good agreement with those obtained by a fixed value of the efficiency with only marginal differences around the end of reionization where the free efficiency prefers a faster end.\\

In our previous work we assumed a conservative truncation magnitude in the UV luminosity function, at -17, that excluded the contribution of the low-luminosity tail of the function which shows an acceleration still under investigation. We have now tested a more optimistic truncation magnitude of -15 showing it has a little impact on the cosmological parameters but prefers a larger derived optical depth of $\tau=0.0541^{+0.0013}_{-0.0016}$ and longer duration of reionization $ \Delta_z^{\rm Reion}=4.22_{-0.62}^{+0.33}$ and it impacts the reconstructed reionization history. The contribution by the faintest sources causes a preference for shallower reionization histories with a larger contribution from the high redshift part with respect to the more conservative truncation. 
The impact of our findings for the reionization history for CMB are displayed in \autoref{SpectPost}, where we show the residuals angular power spectra with respect to Planck hyperbolic tangent best fit of the case considering free efficiency and the case with also the UV more optimistic cut respectively in darker  and lighter colors with respect to the B1 standard case with fixed efficiency. 
These results offer an intriguing perspective for future experiments which will be able to reduce the uncertainty in the faint end of the luminosity function.\\ 

%Considering their capabilities to tighten the constraints on the ionization efficiency strongly reducing the error bars we tested different selections of the QHII data in order to investigate possible differences arising within the whole dataset we used in our previous work. 
Given their constraining power, we tested different selections of QHII data.
The results from the full dataset are robust against different changes of dubious data points and also w.r.t. the exclusion of the only GRB data point we used. Note however that the change in the error bars suggests that GRBs might be an interesting source of constraints from future experiments.
Some differences arise when splitting the QHII between redsfhits lower and higher than 7. The lower dataset has a greater power, as expected from the lower uncertainties on the data points but we note how the higher redshift dataset prefers an higher optical depth $\tau=0.0534\pm 0.0016$
with respect to the lower $\tau=0.0513^{+0.0015}_{-0.0010}$. Also the reconstructed reionization histories are slightly different, the higher redshift dataset seems to prefer an earlier reionization, although with a similar shape and shallowness to the low-z data.\\

The thorough investigation of both theoretical priors and our selection of astrophysical data performed in this paper confirmed our previous results in \cite{PRL}. Astrophysical data are consistent with Planck measurements and provide the necessary leverage to constrain the ionization efficiency, leading to results consistent with the fixed value assumed in \cite{PRL}. The determination of the optical depth is only marginally sensitive to the theoretical priors when CMB is combined with our selection of astrophysical data, consistently with the results of \cite{Qin:2020xrg}, which however used a different selection of astrophysical data and a different methodology. The reconstructed reionization histories are more sensitive to theoretical priors and astrophysical data cuts, the major change occurring when we change the truncation magnitude for the UV luminosity function. 

The results of this paper show how for our selection of data the value of the integrated optical depth as reconstructed in our methodology is mainly driven by astrophysical data and only sightly sensitive to actual history of reionization. Results from very different histories produce similar optical depth although with some shift towards lower values when low redshift quasars are added to the data combinations. UV data are in good agreement with a larger optical depth closer to the Planck 2018 value obtained by a hyperbolic tangent modelling. The other cosmological parameters which are instead mainly constrained by the CMB show very little sensitivity to the change in the optical depth with only a fraction of sigma shift in the scalar spectral index. The little sensitivity of the cosmological parameters to the change in the reionization history may lead to favour a split of the data, reionization dynamics constrained through astrophysical data and the cosmological model based on a simple optical depth prior. As also shown in \cite{Qin:2020xrg}, the differences in the E-mode polarization power spectrum among different reionization reconstructions are below the sensitivity of current data, but this will not be true anymore for future experiments which will reach the cosmic variance limit. At the same time it has been shown in \cite{HPBFSSS,HPFS18} how future experiment will reach a level of sensitivity for which the extended cosmological and primordial power spectrum models will be degenerate with the reionization history and therefore the astrophysical data combination with the CMB will reach a crucial importance in constraining extended cosmological models.

The overall results point to even brighter synergies for the reconstruction of reionization
in the perspectives of future observations. We expect progress in the measurement of the CMB
E-mode polarization on large angular scales from CLASS \citep{Dahal:2019xuf} and LiteBIRD \citep{Sugai:2020pjw}, of the UV luminosity function which will clarify
the faint end slope and finally of the high redshift QHII data either from
Quasars or Gamma Ray Bursts.

%%%%%%%%%%%%%%%%%%%%%%%%%%%%%%%%%%%%%%%%%%%%%%%%%%%%%%%%%%%%%%%%%%%%%%%%%%%%%%%
\begin{acknowledgments}
The authors thank Masami Ouchi and Masafumi Ishigaki for providing their UV luminosity density data compilation.
DP and FF acknowledge financial support by ASI Grant 2016-24-H.0 and the agreement n. 2020-9-HH.0 ASI-UniRM2 ``Partecipazione italiana alla fase A della missione LiteBIRD".
DP acknowledges the computing centre  of Cineca and INAF, under the coordination of the ``Accordo Quadro MoU per lo svolgimento di attività congiunta di ricerca Nuove frontiere in Astrofisica: HPC e Data Exploration di nuova generazione", for the availability of computing resources and support with the project INA17-C5A42. We also acknowledge the use of Cineca under the INFN agreement for indark.
The authors acknowledge  the  usage  of  computational  resources at the Institute of Mathematical Science’s High Performance Computing facility (hpc.imsc.res.in) [Nandadevi].
GFS acknowledges Laboratoire APC-PCCP, 
and also the financial support of the UnivEarthS Labex program at Universite de Paris  (ANR-10-LABX-0023 and ANR-11-IDEX-0005-02). 
\end{acknowledgments}

%%%%%%%%%%%%%%%%%%%%%%%%%%%%%%%%%%%%%%%%%%%%%%%%%%%%%%%%%%%%%%%%%%%%%%%%%%%%%%%
\bibliography{reionref}

\end{document}